\documentclass[11pt,a4paper]{article}
\usepackage[utf8]{inputenc}
\usepackage{times}
\usepackage[T1]{fontenc}
\usepackage[english]{babel}
\usepackage[pdftex]{graphicx}
\usepackage{xcolor}
\usepackage{a4}
\usepackage{amsfonts,amsmath,amssymb,bbm,slashed}
\usepackage[amssymb,Gray]{SIunits}
\usepackage{url}
\usepackage[numbers,sort&compress]{natbib}
\usepackage{booktabs}
\usepackage[a4paper]{geometry}
\usepackage[margin=\parindent,font=small,labelfont=bf]{caption}
\usepackage[caption=false]{subfig}

\renewcommand\appendix{\par
  \setcounter{section}{0}%
  \setcounter{subsection}{0}%
  \setcounter{figure}{0}%
  \renewcommand\thesubsection{\Alph{subsection}}%
  \renewcommand\thefigure{\Alph{subsection}.\arabic{figure}}
  \numberwithin{equation}{subsection}}

\newcommand{\rmi}{\ensuremath{\mathrm{i}}}
\newcommand{\rme}{\ensuremath{\mathrm{e}}}

\newcommand{\erf}{\ensuremath{\mathrm{erf}}}
\newcommand{\abs}[1]{\ensuremath{\left\vert #1 \right\vert}}

\newcommand{\order}[1]{\mathcal{O}(#1)}
\newcommand{\bleq}{\ensuremath{\mathrel{\phantom{=}}}}
\newcommand{\nnl}{\nonumber\\}

\newcommand{\bra}[1]{\langle #1 \hspace{-2pt} \mid}
\newcommand{\ket}[1]{\mid \hspace{-1pt} #1 \rangle}

\newcommand{\D}{\mathrm{d}}
\newcommand{\laplace}{\nabla^2}
\newcommand{\sne}{Schr{\"o}\-din\-ger--New\-ton equation}
\newcommand{\schr}{Schr{\"o}\-din\-ger}

\renewcommand{\vec}[1]{\boldsymbol{\mathrm{#1}}}

\title{Effects of Newtonian gravitational self-interaction
in harmonically trapped quantum systems}
\author{Andr\'e Gro{\ss}ardt\textsuperscript{1,2},
  \quad James Bateman\textsuperscript{3},
  \quad Hendrik Ulbricht\textsuperscript{3},
  \quad Angelo Bassi\textsuperscript{1,2} \\
  \textsuperscript{1}\;\footnotesize\textit{Department of Physics, University of Trieste, 34151 Miramare-Trieste, Italy}\\[-5pt]
  \textsuperscript{2}\;\footnotesize\textit{Istituto Nazionale di Fisica Nucleare, Sezione di Trieste, Via Valerio 2, 34127 Trieste, Italy}\\[-5pt]
  \textsuperscript{3}\;\footnotesize\textit{School of Physics and Astronomy, University of Southampton, SO17 1BJ, United Kingdom}\\
  \footnotesize\textit{Electronic mail addresses: andre.grossardt@ts.infn.it, jbateman@soton.ac.uk,}\\[-8pt]
  \footnotesize\textit{h.ulbricht@soton.ac.uk, bassi@ts.infn.it}
}

\begin{document}

\maketitle

\begin{abstract}\noindent
The \sne\ has gained attention in the recent past as a nonlinear modification of the \schr\ equation due to a
gravitational self-interaction. Such a modification is expected from a fundamentally semi-classical theory of
gravity, and can therefore be considered a test case for the necessity of the quantisation of the gravitational
field. Here we provide a thorough study of the effects of the \sne\ for a micron-sized sphere trapped in a harmonic
oscillator potential. We discuss both the effect on the energy eigenstates and the dynamical behaviour of squeezed
states, covering the experimentally relevant parameter regimes.
\end{abstract}

\section{Introduction}

The interaction of nonrelativistic quantum matter with an \emph{external} gravitational field has been experimentally
established by the famous COW experiment~\cite{Colella:1975}. Be that as it may, the question
whether gravity is fundamentally a quantum theory resembling the other fields, or something different, is still open.
There is no unambiguous answer to the question how quantum matter sources the gravitational field.
While the standard approach in regard to the great success of quantum field theory is to quantise the gravitational
field along similar lines, there is no experimental evidence, nor a strict theoretical necessity, to date, that
the gravitational field must be quantised~\cite{Rosenfeld:1963,Mattingly:2005,Kiefer:2007}.

Taking the possibility of a fundamentally classical description of space-time into account, the most natural way
to describe the interaction of quantum matter with such a classical space-time within the framework of general
relativity is provided by the semi-classical Einstein equations,
\begin{equation}
\label{eqn:sce}
R_{\mu \nu} + \frac{1}{2} g_{\mu \nu} R = \frac{8 \pi G}{c^4} \,
\bra{\Psi} \hat{T}_{\mu \nu} \ket{\Psi} \,,
\end{equation}
i.\,e. Einstein's field equations where the energy-momentum tensor is replaced by the expectation value of a
corresponding quantum operator in some quantum state $\Psi$; a theory that was already suggested in the
1960s by Møller~\cite{Moller:1962} and Rosenfeld~\cite{Rosenfeld:1963}.\footnote{Although
it is often claimed that a fundamentally semi-classical theory of gravity was ruled out by
experiment~\cite{Page:1981}, the arguments against such a theory are inconclusive; cf. the discussions
in references~\cite{Mattingly:2005,Kiefer:2007,Bahrami:2014}. It needs to be stressed that, different than in other
situations where semi-classical gravity is considered as an effective limit of some underlying quantum theory
of gravity~\cite{Anastopoulos:2014a}, in this approach equation~\eqref{eqn:sce} is taken as fundamental.}

In the nonrelativistic limit, such a fundamentally semi-classical theory of gravity adds a nonlinear potential
term to the \schr\ equation~\cite{Giulini:2012,Bahrami:2014}. The resulting equation is known as the \emph{\sne}.
For a multi-particle system, it reads
\begin{subequations}\label{eqn:n-particle-sn}\begin{align}
\rmi \hbar \frac{\partial}{\partial t} \Psi(t,\vec r_1,\cdots,\vec r_N)
&= \biggl(-\sum_{i=1}^N\frac{\hbar^2}{2m_i}\laplace_i + V_\text{ext} +  V_g[\Psi] \biggr)\Psi(t;\vec r_1,\cdots \vec r_N) \\
V_g[\Psi](t,\vec r_1,\cdots,\vec r_N) &= -G\sum_{i=1}^N\sum_{j=1}^N m_i m_j \int \D^3 r_1' \cdots \D^3 r_N'
 \frac{\abs{\Psi(t;\vec r'_1,\cdots,\vec r'_N)}^2}{\abs{\vec r_i - \vec r'_j}} \,,
\end{align}\end{subequations}
where $\Psi$ is the $N$-particle wave-function, and $V_g$ is the gravitational interaction. The \sne\ becomes
nonlinear due to the dependence of $V_g$ on the absolute-value squared of the wave-function.
An intuitive way of looking at this equation is that the probability density, $\abs{\Psi}^2$, acts like a mass density
generating a Newtonian gravitational potential, which then appears in the \schr\ equation in the usual way.
$V_\text{ext}$ is an external, linear potential, which will be a quadratic, i.\,e. harmonic oscillator, potential
here.

Equation \eqref{eqn:n-particle-sn} was first considered by Diósi~\cite{Diosi:1984} as a model for wave-function 
localisation. Because of its derivability from
semi-classical gravity, it was suggested that the \sne\ can provide evidence for or against the necessity to
quantise the gravitational field~\cite{Carlip:2008}. The original subject of such a test were
heavy molecules in interferometry experiments~\cite{Arndt:2005} for which the \sne\ predicts inhibitions
of the dispersion of the centre-of-mass
wave-function~\cite{Giulini:2011,Meter:2011,Giulini:2013,Giulini:2014,Colin:2014}.
Although the parameter regime where this effect shows up is much closer to the scope of current experiments
than any quantum gravity effect studied so far, the required masses are still several orders of magnitude above
what is currently feasible.

An alternative test of the \sne, using macroscopic quantum systems in a harmonic trap potential, was given
by Yang et\,al.~\cite{Yang:2013}, where it has been shown that the \schr--Newton dynamics lead to a phase difference
between the external and internal oscillations of a squeezed Gaussian state.
Here, we complement this proposal by a more general analysis of effects of the \sne\ on harmonically trapped quantum
systems, going beyond the limit of narrow wave-functions and considering also the regime where the width of the
wave-function becomes comparable to the localisation length of the atoms in the considered microsphere.
In addition to the dynamical effects, we also discuss the gravitational perturbation of the spectrum of the stationary
energy eigenstates.

While we will find that the dynamical effect on the internal structure of a squeezed state is indeed strongest in
the limit of a narrow wave-function, as it has been studied by Yang et\,al.~\cite{Yang:2013}, the intermediate regime
is the most suitable to observe effects in the energy spectrum. These turn out to be of comparable order of magnitude
as the dynamical effects. However, their observation requires slightly smaller masses and, more importantly,
there is no necessity to create a squeezed state, nor for quantum state tomography, which makes an observation
more feasible.

We present the Hamiltonian for the trapped system with Newtonian self-gravitational interaction in the second
section. We derive an approximation for the gravitational interaction inside a crystalline, or solid, spherical
many-particle system and discuss the reduction of the three-dimensional equation to a one-dimensional \sne, which is
the basis for the discussion thereafter.
In the third section, we study the effects of the \schr--Newton interaction on the energy spectrum. We discuss the
limiting cases of a narrow and wide wave-function, as well as the intermediate regime.
In the fourth section, the dynamical behaviour of a squeezed Gaussian state is derived, recovering the results
from reference~\cite{Yang:2013}. Their results are extended to the regime of finite, non-narrow wave-function sizes.
Finally, our results and the prospects for experimental tests of the \sne\ are summarised in the Conclusions section.

\section{Hamiltonian of a self-gravitating trapped sphere}

Consider a three-dimensional Hamiltonian of a self-gravitating quantum system in an external potential:
\begin{equation}\label{eqn:hamiltonian}
 H = \frac{\vec p^2}{2 m} + V_\text{ext}(\vec r) + V_g[\psi](t,\vec r) \,.
\end{equation}
The coordinates are written as $\vec r = (x,y,z)$. We will specify the external potential $V_\text{ext}$ later.

The Hamiltonian \eqref{eqn:hamiltonian} is supposed to describe the centre-of-mass of a many-particle system.
The gravitational potential, which is a function of all particle coordinates, does, however, not separate into
centre-of-mass and relative coordinates exactly. Such a separation can only be achieved within a suitable
Born--Oppenheimer-type approximation, as has been demonstrated in reference~\cite{Giulini:2014}. The multi-particle
gravitational potential can then be reduced to
\begin{subequations}\label{eqn:vg-born-oppen}\begin{align}
V_g[\psi](t,\vec r) &= -G \, \int \D^3 r' \, \abs{\psi(t,\vec r')}^2 \, I_{\rho_c}(\vec{r} - \vec r')  \\
I_{\rho_c}(\vec d) &= \int \D^3 x \, \D^3 y \, \frac{\rho_c(\vec x) \, \rho_c(\vec y - \vec d)}{\abs{\vec x - \vec y}} \,,
\label{eqn:grav-self-interaction}
\end{align}\end{subequations}
where $\psi$ is the centre-of-mass wave-function, $\vec r$ is the centre-of-mass coordinate, and $\rho_c$ is
the mass density relative to the centre of mass. For a lump of matter, i.\,e. a molecule, of $N$ atoms which is
described by a stationary relative wave-function $\chi$, $\rho_c$ is given as the sum of the marginal
distributions for all but one\footnote{The distribution of the $N$-th particle is given by the centre-of-mass
wave-function and can be neglected if $N$ is sufficiently large.} atoms:
\begin{equation}\begin{split}
\rho_c(\vec r)
=\sum_{i=1}^{N-1} m_i \int \D^3 r_1 \cdots \D^3 r_{i-1} \, \D^3 r_{i+1} \cdots \D^3 r_{N-1} \\
\times \abs{\chi(\vec r_1,\dots,\vec r_{i-1},\vec r,\vec r_{i+1},\dots,\vec r_{N-1})}^2\,.
\end{split}\end{equation}
$G\,I_{\rho_c}$ is simply the gravitational potential energy
between the mass distribution described by $\rho_c$ and the same mass distribution, shifted by $\vec d$.
For a homogeneous, spherical mass distribution with radius $R$ it is given by~\cite{Iwe:1982}
\begin{equation}\label{eqn:self-energy-homogeneous-sphere}
I_{\rho_c}^\text{sphere}(d)=\frac{m^2}{R}\times
\begin{cases}
\frac{6}{5}-2\left(\frac{d}{2R}\right)^2+\frac{3}{2}\left(\frac{d}{2R}\right)^3-\frac{1}{5}\left(\frac{d}{2R}\right)^5
&\text{for}\ d\leq 2R\,,\\
\frac{R}{d}
&\text{for}\ d\geq 2R\,.
\end{cases}
\end{equation} 

Given a solution $\psi^{(0)}$ of the free \schr\ equation (without the gravitational potential $V_g$),
switching on the state dependent gravitational potential~\eqref{eqn:vg-born-oppen} will distort both the energy
expectation value and the shape of the solution.
To first order in the gravitational constant $G$, the correction to the \schr\ evolution due to the nonlinear
\schr--Newton gravitational potential term can be obtained in a perturbation expansion.
For this purpose, we make the ansatz
\begin{equation}
\psi(t,\vec r) = \psi^{(0)}(t,\vec r) + G \, \psi^{(1)}(t,\vec r) + \mathcal{O}(G^2)
\end{equation}
for the wave-function.
Now note that the perturbation $V_g$ of the Hamiltonian can be expanded as
\begin{equation}\label{eqn:Vgpert}
V_g[\psi](t,\vec r) = V_g[\psi^{(0)}](t,\vec r) + \mathcal{O}(G^2) \,,
\end{equation}
where the first term is already $\mathcal{O}(G)$. $V_g[\psi^{(0)}]$ is just a linear correction to the Hamiltonian,
which is time-independent for a stationary state $\psi^{(0)}$. The Hamiltonian~\eqref{eqn:hamiltonian} then takes
the linear form
\begin{equation}\label{eqn:hamiltonian-pert}
 H = \frac{\vec p^2}{2 m} + V_\text{ext}(\vec r) + V_g[\psi^{(0)}](t,\vec r) + \order{G^2} \,.
\end{equation}

This is a good approximation as long as the gravitational interaction is considered to be
weak, and therefore the difference in the wave-function between the solutions of the
unperturbed \schr\ equation and those of the full \sne\ is small.

The potential~\eqref{eqn:vg-born-oppen} can be significantly simplified in the limits where the wave-function is
very narrow or very wide.
Provided that the spatial centre-of-mass wave-function is wide compared to the extent of
the considered many-particle system, the mass distribution within the system plays no significant role, and
the gravitational potential is approximately the same as in the one-particle case, namely~\cite{Giulini:2014}
\begin{equation}
 V_g^\text{wide}[\psi](t,\vec r)
= - G\,m^2 \, \int \D^3 r' \, \frac{\abs{\psi(t,\vec r')}^2}{\abs{\vec r - \vec r'}} \,.
\end{equation}

Consider, on the other hand, the case where the spatial centre-of-mass wave-function
is narrow compared to the extent of the many-particle system, or---to be more precise---where
$I_{\rho_c}$ does not vary too much over the width of the centre-of-mass wave-function.
In this case the potential~\eqref{eqn:vg-born-oppen} can be expanded in a Taylor series in
$\abs{\vec r - \vec r'}$ up to second order, yielding~\cite{Yang:2013,Giulini:2014}
\begin{equation}\label{eqn:narrow-wf-limit}
 V_g^\text{narrow}[\psi](t,\vec r) = -G\,I_{\rho_c}(\vec 0) - \frac{G}{2} I''_{\rho_c}(\vec 0)  \left( \vec r^2
- 2 \,\vec r \cdot \bra{\psi} \vec r \ket{\psi} + \bra{\psi} \vec r^2 \ket{\psi} \right) \,,
\end{equation}
$I''_{\rho_c}$ denoting the Hessian of $I_{\rho_c}$.

\begin{figure}
\centering
\includegraphics[scale=.67]{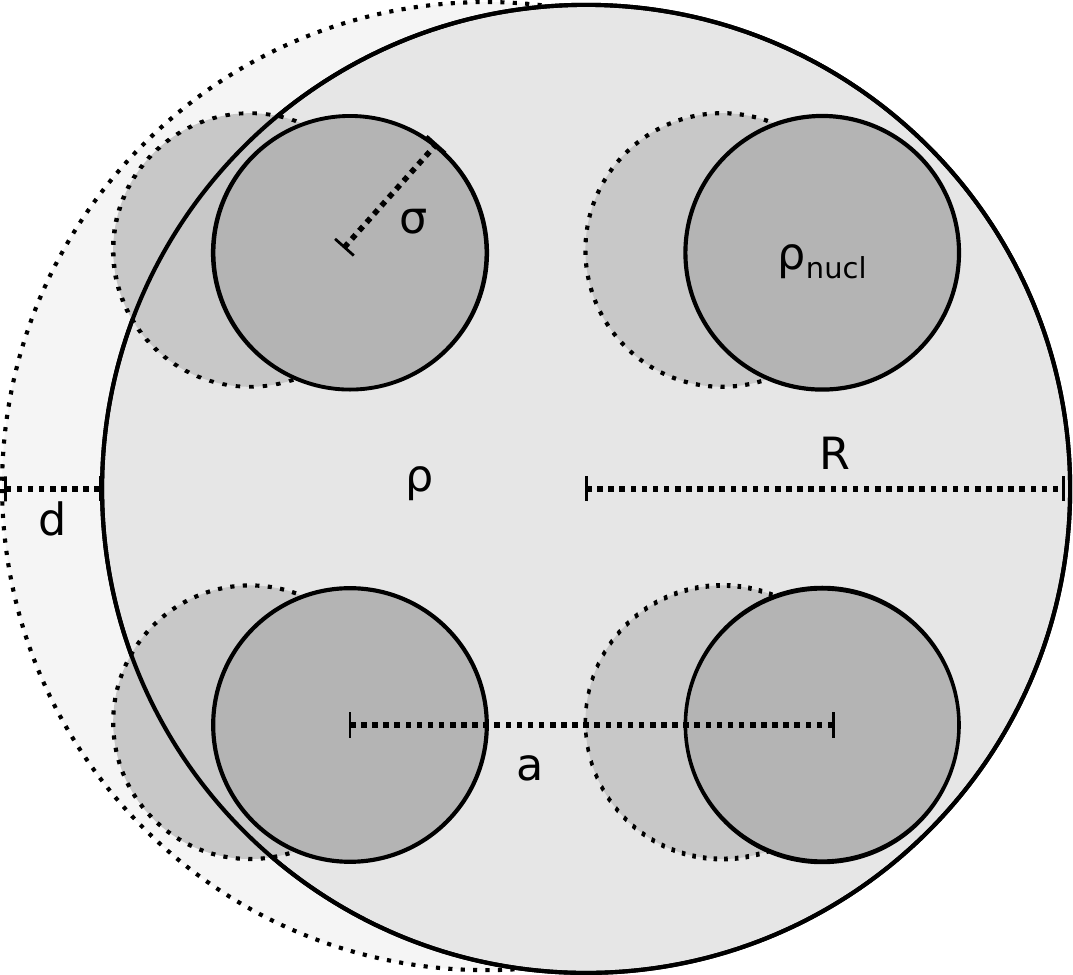} 
\caption{Schematic picture of how the function $I_{\rho_c}$ is determined for a sphere of
atoms in a cubic lattice.}
\label{fig:lattice}
\end{figure}

If the mass is assumed to be distributed homogeneously over a sphere of radius $R$, and therefore the function
$I_{\rho_c}$ takes the form~\eqref{eqn:self-energy-homogeneous-sphere},
then the potential is~\cite{Giulini:2014}
\begin{equation}\label{eqn:grav-pot-sphere}
 V^\text{narrow, sphere}_g[\psi](t,\vec r) = - \frac{G \, m^2}{R^3} \left( \frac{6}{5} \, R^2 - \frac{\vec r^2}{2}
+ \vec r \cdot \bra{\psi} \vec r \ket{\psi}
-\frac{\bra{\psi} \vec r^2 \ket{\psi}}{2} \right) \,.
\end{equation}

However, as pointed out by Yang et\,al.~\cite{Yang:2013}, a realistic microsphere has a crystalline substructure,
which must be taken into account if the wave-function is narrow enough to probe the atomic regime.

\subsection{Crystalline substructure}

A more realistic mass distribution should account for the fact that most of the mass in a crystalline
structure is well-localised around the positions of the nuclei. $I_{\rho_c}(\vec d)$ then
represents the gravitational interaction of a grid of $N$ atoms with an identical grid, shifted by distance~$\vec d$.
We model the quantum system as a sphere of radius $R$, within which the atoms are homogeneous spheres of
radius $\sigma$, as depicted in figure~\ref{fig:lattice}. There are two contributions to $I_{\rho_c}(\vec d)$:
\begin{enumerate}
\item The self-energy of each atom with its own ``copy'' which, approximately, can be modelled as the gravitational
self-interaction of a sphere of radius $\sigma$ with mass $m/N$, hence
\begin{equation}
I_\text{self}(\vec d) = N \times \left(\frac{(m/N)^2}{\sigma}\right) \times
\begin{cases}
\frac{6}{5}-2\left(\frac{d}{2 \sigma}\right)^2+\frac{3}{2}\left(\frac{d}{2 \sigma}\right)^3-\frac{1}{5}\left(\frac{d}{2 \sigma}\right)^5
&\text{for}\ d\leq 2 \sigma\,,\\
\frac{\sigma}{d}
&\text{for}\ d\geq 2 \sigma\,.
\end{cases}
\end{equation}
\item The mutual interaction of each atom with all $N-1$ other atoms, which is the Riemann sum for the
integral~\eqref{eqn:grav-self-interaction} for the full sphere of radius $R$, if the sphere is split into
$N$ sub-areas of volume $a^3$, hence
\begin{align}
I_\text{mutual}(\vec d) &= \sum_{i=1}^N \sum_{j=1 \atop j\not=i}^N
\left( \frac{ (m/N)^2}{\abs{\vec r_i - \vec r_j - \vec d}}\right) \nnl
&= \int \D^3 x \, \D^3 y
\frac{\rho_\text{sphere}(\vec x) \, \rho_\text{sphere}(\vec y - \vec d)}{\abs{\vec x - \vec y}}
+ \mathcal{O}\left(N^{-5/3}\right) \nnl
&= \frac{m^2}{R} \times
\begin{cases}
\frac{6}{5}-2\left(\frac{d}{2R}\right)^2+\frac{3}{2}\left(\frac{d}{2R}\right)^3-\frac{1}{5}\left(\frac{d}{2R}\right)^5
&\text{for}\ d\leq 2R\,,\\
\frac{R}{d}
&\text{for}\ d\geq 2R\,.
\end{cases}
\end{align}
\end{enumerate}
Therefore, for large $N$, the total function  $I_{\rho_c}$ for a crystalline sphere is
\begin{subequations}\begin{align}\label{eqn:I-lattice}
I_\text{cr}^\text{sphere}(\vec d) &= I_\text{self}(\vec d) + I_\text{mutual}(\vec d) \nnl
&= \frac{m^2}{N \, \sigma} \times
\begin{cases}
\frac{6}{5} \gamma_0 - 2 \gamma_2 \left(\frac{d}{2 \sigma}\right)^2
+\frac{3}{2} \gamma_3 \left(\frac{d}{2 \sigma}\right)^3-\frac{1}{5} \gamma_5 \left(\frac{d}{2 \sigma}\right)^5
&\text{for}\ d\leq 2 \sigma\,,\\
\frac{\sigma}{d} + \frac{6}{5} \beta_0 - 2 \beta_2 \left(\frac{d}{2 \sigma}\right)^2
+\frac{3}{2} \beta_3 \left(\frac{d}{2 \sigma}\right)^3-\frac{1}{5} \beta_5 \left(\frac{d}{2 \sigma}\right)^5
&\text{for}\ d\leq 2R\,,\\
(N+1) \, \frac{\sigma}{d}
&\text{for}\ d\geq 2R\,,
\end{cases}
\intertext{with}
\gamma_k &= 1 + \beta_k \,; \quad\quad\quad
\beta_k = N \, \left(\frac{\sigma}{R}\right)^{k+1}
= \frac{\sigma^{k+1}}{a^3\,R^{k-2}} \,.
\end{align}\end{subequations}
Note that the expansion~\eqref{eqn:narrow-wf-limit} is still valid in the limit of a narrow wave-function,
which now means that the width of the wave-function is small compared to the atomic radius $\sigma$.
Making use of the fact that $\gamma_2 \approx 1$ for $R \gg a \gg \sigma$, the corresponding
gravitational potential is
\begin{subequations}\begin{align}\label{eqn:grav-pot-lattice}
 V^\text{cr}_g[\psi](t,\vec r) &= m \, \omega_\text{SN}^2
\,\left( -\frac{6}{5} \, \gamma_0 \, \sigma^2
+ \frac{\vec r^2}{2}
- \vec r \cdot \bra{\psi} \vec r \ket{\psi}
+ \frac{\bra{\psi} \vec r^2 \ket{\psi}}{2} \right)
\intertext{with}
\omega_\text{SN}^\text{sphere} &= \sqrt{\frac{G\,m_\text{atom}}{\sigma^3}}\,,
\label{eqn:omega-sn-sphere}
\end{align}\end{subequations}
where $m_\text{atom}$ is the mass of a single atom in the crystal.
This potential has been used in reference~\cite{Yang:2013} to describe the behaviour of a narrow squeezed coherent
state in a harmonic trap. Since it is quadratic in $\vec r$, a Gaussian state will remain
Gaussian~\cite{Yang:2013,Colin:2014}, but there will be a gravitational contribution to the
coupling constant. We come back to this in section~\ref{sec:dynamics}.

If the atoms are, more realistically, modelled by Gaussian matter distributions (cf. \cite{Yang:2013}),
\begin{equation}
\rho_c^\text{Gauss}(\vec x)
= \frac{m}{\sqrt{\pi^3}\,N\,\sigma^3} \, \exp\left(-\frac{x^2}{\sigma^2}\right) \,,
\end{equation}
one can see with equation~\eqref{eqn:grav-self-interaction} that the self-interaction part of the
function $I_\text{cr}$ takes the form~\cite{Iwe:1982}
\begin{equation}
I_\text{self}^\text{Gauss}(\vec d) = \frac{m^2}{N\,d} \, \erf\left( \frac{d}{\sqrt{2}\,\sigma} \right)\,,
\end{equation}
where $\erf$ is the Gauss error function. The total $I_\text{cr}$ then is
\begin{align}\label{eqn:I-lattice-exp}
I_\text{cr}^\text{Gauss}(\vec d) &= \frac{m^2}{N \, \sigma} \times \nnl
&\bleq \hspace{-1cm} \times \begin{cases}
\frac{\sigma}{d}\,\erf\left( \frac{d}{\sqrt{2}\,\sigma} \right) + \frac{6}{5} \beta_0 - 2 \beta_2 \left(\frac{d}{2 \sigma}\right)^2
+\frac{3}{2} \beta_3 \left(\frac{d}{2 \sigma}\right)^3-\frac{1}{5} \beta_5 \left(\frac{d}{2 \sigma}\right)^5
&\text{for}\ d\leq 2R\,,\\
\left(N+\erf\left( \frac{d}{\sqrt{2}\,\sigma} \right)\right) \, \frac{\sigma}{d}
&\text{for}\ d\geq 2R\,.
\end{cases}
\end{align}
For the Gaussian matter distribution, the gravitational potential of a narrow wave-function is of the same
shape~\eqref{eqn:grav-pot-lattice}, but with the frequency $\omega_\text{SN}$ and $\gamma_0$ replaced
by\footnote{This result should in principle agree with equation (14) of reference~\cite{Yang:2013},
provided that their $\Delta x_\text{zp} = \sqrt{\langle x^2 \rangle} = \sigma/\sqrt{2}$.
However, we find a factor of $\sqrt{2}$ difference compared to their expression for $\omega_\text{SN}^\text{Gauss}$.}
\begin{equation}\label{eqn:omega-sn-gauss}
\omega_\text{SN}^\text{Gauss} = \sqrt{\sqrt{\frac{2}{\pi}}\,\frac{G \, m_\text{atom}}{3\,\sigma^3}}\,,
\quad\quad\quad
\gamma_0^\text{Gauss} = \frac{5}{2} + \frac{3}{2} \, \sqrt{2 \pi} \, \beta_0 \,,
\end{equation}
where we assumed $R^3 \gg N\,\sigma^3$.

It should be remarked that, while the splitting in cubes of volume $a^3$ in the derivation provided here seems to
imply the requirement of a simple cubic crystal structure, the result is actually independent of the type of the
present crystal structure. Even a non-crystalline, amorphous substructure will still exhibit the behaviour described
here, as long as the localisation length $\sigma$ of the atoms is small compared to the average distance $a$ between
the atoms.

\subsection{Reduction to one dimension}\label{sec:reduction}

An approximate one-dimensional version of the \sne\ can be obtained in the case where the shape of the external
potential is such that the wave-function will be narrow in the two remaining dimensions. In this case, where the
wave-function satisfies approximately
\begin{equation}\label{eqn:axial-wf}
\abs{\psi(x,y,z)}^2 = \abs{\psi(x)}^2 \, \delta(y) \, \delta(z)\,,
\end{equation}
the gravitational potential~\eqref{eqn:vg-born-oppen} takes the form
\begin{equation} \label{eqn:vg-one-dim}
V_g[\psi](t,x) = -G \, \int \D x' \, \abs{\psi(x')}^2 \, I_{\rho_c}(\abs{x-x'}) \,.
\end{equation}
$I_{\rho_c}$ is an even function by definition for any matter distribution $\rho_c$, hence the absolute value in
the argument of $I_{\rho_c}$. The dependence of the argument on $y$ and $z$ can be neglected, because the parts
where $y$ or $z$ is significantly different from zero do not contribute in the \schr\ equation after
multiplication with the wave-function.
Substituting $d = \abs{x-x'}$ the potential can be rewritten as
\begin{equation}\label{eqn:axial-grav-potential}
V_g[\psi](t,x) = -G \, \int_0^\infty \D d \, \left( \abs{\psi(x-d)}^2 + \abs{\psi(x+d)}^2 \right) \, I_{\rho_c}(d) \,.
\end{equation}

The functions~\eqref{eqn:I-lattice} and~\eqref{eqn:I-lattice-exp} can now be applied to this one-dimensional
potential without any changes, and the \schr\ equation separates and yields the one-dimensional equation
\begin{equation}\label{eqn:sn-1d}
\rmi\hbar\,\frac{\partial}{\partial t} \psi(t,x) = -\frac{\hbar^2}{2\,m}\,\frac{\partial^2}{\partial x^2} \psi(t,x)
+ \frac{m}{2}\, \omega_0^2\, x^2\,\psi(t,x) + V_g[\psi]\,\psi(t,x)\,.
\end{equation}
Note that we assume now that the external potential is quadratic with trap frequency $\omega_0$ in $x$-direction,
while the shape of the external potential in $y$- and $z$-direction does not play a role, as long as the
wave-function will be narrow.

The one-dimensional potential~\eqref{eqn:axial-grav-potential} still has the corresponding limits
\begin{subequations}\begin{align}
V_g^\text{wide}[\psi](t,x)
&= - G\,m^2 \, \int \D x' \, \frac{\abs{\psi(t,x')}^2}{\abs{x - x'}}
\label{eqn:axial-wide-limit}\\
V_g^\text{narrow}[\psi](t,x) &= -G\,I_{\rho_c}(0) - \frac{G}{2} I''_{\rho_c}(0)  \left( x^2
- 2 \,x \, \langle x \rangle + \langle x^2 \rangle \right)
\label{eqn:axial-narrow-limit}
\end{align}\end{subequations}
for a wide and narrow wave-function, respectively.

\section{Gravitational effects on the energy spectrum}\label{sec:spectrum}

Without the gravitational potential $V_g$, the \schr\ equation~\eqref{eqn:sn-1d} has the well known
energy eigenstates
\begin{subequations}\begin{align}
\label{eqn:harm-osci-states}
\psi^{(0)}_{n}(x) &= \frac{1}{\sqrt{2^n \, n!}} \, \left( \frac{m\,\omega_0}{\pi\,\hbar} \right)^{1/4} \,
\exp \left(-\frac{m\,\omega_0\,x^2}{2 \hbar}\right) \,
H_n\left( \sqrt{\frac{m\,\omega_0}{\hbar}}\,x \right) \,,
\intertext{where the Hermite polynomials $H_n$ are defined by}
\label{eqn:hermite}
H_n(x) &= (-1)^n\, \rme^{x^2} \frac{\D^n}{\D x^n} \rme^{-x^2}
\intertext{and the corresponding energy eigenvalues are}
\label{eqn:unperturbed-energy}
E^{(0)}_{n} &= \hbar \omega_0 \left( \frac{1}{2} + n\right)\,.
\end{align}\end{subequations}
As long as one is only concerned with stationary solutions, one can perform a first-order
perturbation calculation to obtain the energy correction coming from the gravitational potential.
In the quadratic narrow wave-function approximation~\eqref{eqn:grav-pot-lattice} we immediately get the
energy correction
\begin{align}\label{eqn:energy-correction-narrow}
 \Delta E_n &= \bra{\psi^{(0)}_n} V_g[\psi^{(0)}_n](\vec r) \ket{\psi^{(0)}_n} \nnl
 &= m\,\omega_\text{SN}^2 \left( -\frac{6}{5}\,\gamma_0\,\sigma^2
+ \left(n+\frac{1}{2}\right)\,\frac{\hbar}{m\,\omega_0} \right) \,.
\end{align}
In this approximation, the first term is just a constant shift of all energy levels, while the second term
changes the spectral transition energies proportionally to $\omega_\text{SN}^2$. The transition energies,
\begin{equation}\label{eqn:transition-energies-quadratic}
E^\text{trans}_{n_1 n_2} = E_{n_2} - E_{n_1}
= \hbar\,\omega_0\,(n_2 - n_1)\, \left(1 + \frac{\omega_\text{SN}^2}{\omega_0^2} \right)\,,
\end{equation}
are, however, still degenerate, i.\,e. they depend only on the difference $(n_2 - n_1)$, and not on $n_1$ and $n_2$
alone. This degeneracy is removed if the higher order terms in the gravitational potential are taken into account,
leading to a fine-structure of the spectral lines.\footnote{Note that this fine-structure of the harmonic oscillator
is of a different nature than the well-known fine-structure of atomic spectra. While in the latter there is a degeneracy
of the actual energy eigenvalues, that is removed by additional interaction terms, the one-dimensional harmonic
oscillator has an infinite number of \emph{non-degenerate} energy eigenstates whose energy eigenvalues are shifted
due to the \schr--Newton potential. The degeneracy here is in the transition spectrum, where transition energies
between eigenstates depend on the difference $(n_2-n_1)$ only. This is the degeneracy that is removed by the
\schr--Newton term.}

To arrive at equation~\eqref{eqn:energy-correction-narrow} we made use of the approximation~\eqref{eqn:Vgpert}.
As mentioned before, in this case the gravitational potential is just a linear correction and the energy shift
can be calculated in ordinary perturbation theory.
Maintaining this approximation, but now using the full gravitational potential~\eqref{eqn:vg-one-dim} instead of
the quadratic approximation for narrow wave-functions, one obtains:
\begin{align}\label{eqn:energy-correction}
\Delta E_n &= -\frac{G}{(2^n \, n!)^2} \, \frac{m\,\omega_0}{\pi\,\hbar} \,
\int_{-\infty}^\infty \D x \, \exp \left(-\frac{m\,\omega_0\,x^2}{\hbar}\right) \,
H_n\left( \sqrt{\frac{m\,\omega_0}{\hbar}}\,x \right)^2 \nnl
&\bleq \times \int_0^\infty \D d \, 
\Bigg[ \exp \left(-\frac{m\,\omega_0\,(x-d)^2}{\hbar}\right) \,
H_n\left( \sqrt{\frac{m\,\omega_0}{\hbar}}\,(x-d) \right)^2 \nnl
&\bleq + \exp \left(-\frac{m\,\omega_0\,(x+d)^2}{\hbar}\right) \,
H_n\left( \sqrt{\frac{m\,\omega_0}{\hbar}}\,(x+d) \right)^2 \Bigg] \,  I_{\rho_c}(d) \,.
\end{align}
Introducing the dimensionless variables
\begin{equation}
\xi = \sqrt{\frac{m\,\omega_0}{\hbar}}\,x \,, \quad\quad
\zeta = \frac{d}{2\,\sigma} \,, \quad\quad
\alpha = 2\,\sigma\,\sqrt{\frac{m\,\omega_0}{\hbar}} \,, \quad\quad
\varrho = \frac{R}{\sigma} = \sqrt[3]{\frac{3\,m}{4\pi\,\rho\,\sigma^3}} \,,
\end{equation}
we get
\begin{subequations}\begin{align}\label{eqn:energy-correction-simplified}
 \Delta E_n &= -\frac{G\,m_\text{atom}\,m}{\sigma} \, \frac{f_n(\alpha,\,\varrho)}{\alpha^2}
= -\frac{G\,\hbar\,m_\text{atom}}{4\,\sigma^3\,\omega_0} \, f_n(\alpha,\,\varrho)
\intertext{with}
f_n(\alpha,\,\varrho) &= \alpha^3\,\sqrt{\frac{2}{\pi}} \, \int_0^\infty \D \zeta \,
\exp \left(-\frac{\alpha^2\,\zeta^2}{2}\right)\, P_n(\alpha\,\zeta)\,
i(\zeta,\,\varrho)  \,, \label{eqn:energy-correction-fn}
\intertext{the even polynomials}
P_n(z) &= \frac{1}{\sqrt{2\pi}\,(2^n \, n!)^2} \, \exp\left(-\frac{z^2}{2}\right)
 \int_{-\infty}^\infty \D \xi \, \exp \left(-2\xi^2\right) \,H_n\left( \xi \right)^2  \nnl
&\bleq \times \Bigg[ \exp \left(2\,z\,\xi\right) \, H_n\left( \xi - z \right)^2
+ \exp \left(-2\,z\,\xi\right) \, H_n\left( \xi + z \right)^2 \Bigg] \,,
\intertext{and the matter distribution functions}
i_\text{cr}^\text{sphere}(\zeta,\,\varrho) &= \begin{cases}
\frac{6}{5} \gamma_0 - 2 \gamma_2 \zeta^2 +\frac{3}{2} \gamma_3 \zeta^3-\frac{1}{5} \gamma_5 \zeta^5
&\text{for}\ \zeta\leq 1\,,\\
\frac{1}{2\,\zeta} + \frac{6}{5} \beta_0 - 2 \beta_2 \zeta^2 +\frac{3}{2} \beta_3 \zeta^3
-\frac{1}{5} \beta_5 \zeta^5 &\text{for}\ \zeta\leq \varrho\,,\\
\frac{N+1}{2\,\zeta} &\text{for}\ \zeta \geq \varrho
\end{cases}
\intertext{for spherical atoms, and}
i_\text{cr}^\text{Gauss}(\zeta,\,\varrho) &= \begin{cases}
\frac{1}{2\,\zeta}\,\erf\left(\sqrt{2}\,\zeta\right)
+ \frac{6}{5} \beta_0 - 2 \beta_2 \zeta^2 +\frac{3}{2} \beta_3 \zeta^3
-\frac{1}{5} \beta_5 \zeta^5 &\text{for}\ \zeta\leq \varrho\,,\\
\frac{N+\erf\left(\sqrt{2}\,\zeta\right)}{2\,\zeta} &\text{for}\ \zeta \geq \varrho
\end{cases}
\end{align}\end{subequations}
for a Gaussian distribution of the atomic matter density, respectively.
The polynomials $P_n$ can be solved analytically, and so can the functions $f_n$, at least in
principle.\footnote{We obtained $P_n$ for small $n$ up to $n=14$ using \emph{Mathematica}. The time
needed for the evaluation increases, however, exponentially with $n$. $f_n$ can be integrated analytically as
well, in terms of integral functions. For realistic physical parameters it is nevertheless necessary to omit the
constant contribution $f_n^{(0)}$ (see text below for definition) in a numerical evaluation of the transition energies,
since otherwise the subtraction of two comparable, large numbers from each other would lead to a significant loss of
numerical accuracy.} The transition energies can then be calculated as
\begin{subequations}\begin{align}\label{eqn:transition-energies}
E^\text{trans}_{n_1 n_2} &= \hbar\,\omega_0\, \left(n_2 - n_1
- \frac{G\,m_\text{atom}}{4\,\sigma^3\,\omega_0^2}\, \tilde{f}_{n_1 n_2}(\alpha,\varrho) \right)\,, \\
\tilde{f}_{n_1 n_2}(\alpha,\varrho) &= f_{n_2}(\alpha,\varrho) - f_{n_1}(\alpha,\varrho)\,.
\end{align}\end{subequations}

\begin{figure}
\centering
\subfloat[spherical mass distribution, $n=0$]{\includegraphics[scale=.5]{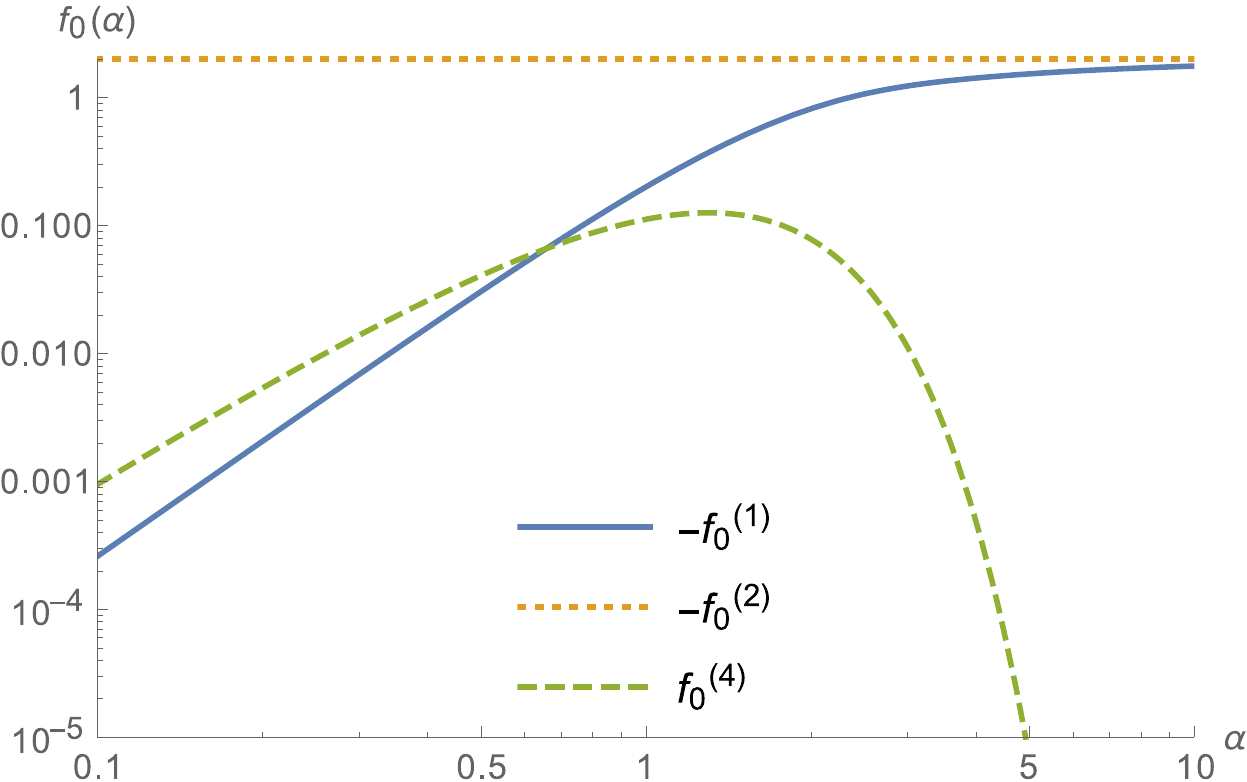}}
\hspace{1cm}
\subfloat[Gaussian mass distribution, $n=0$]{\includegraphics[scale=.5]{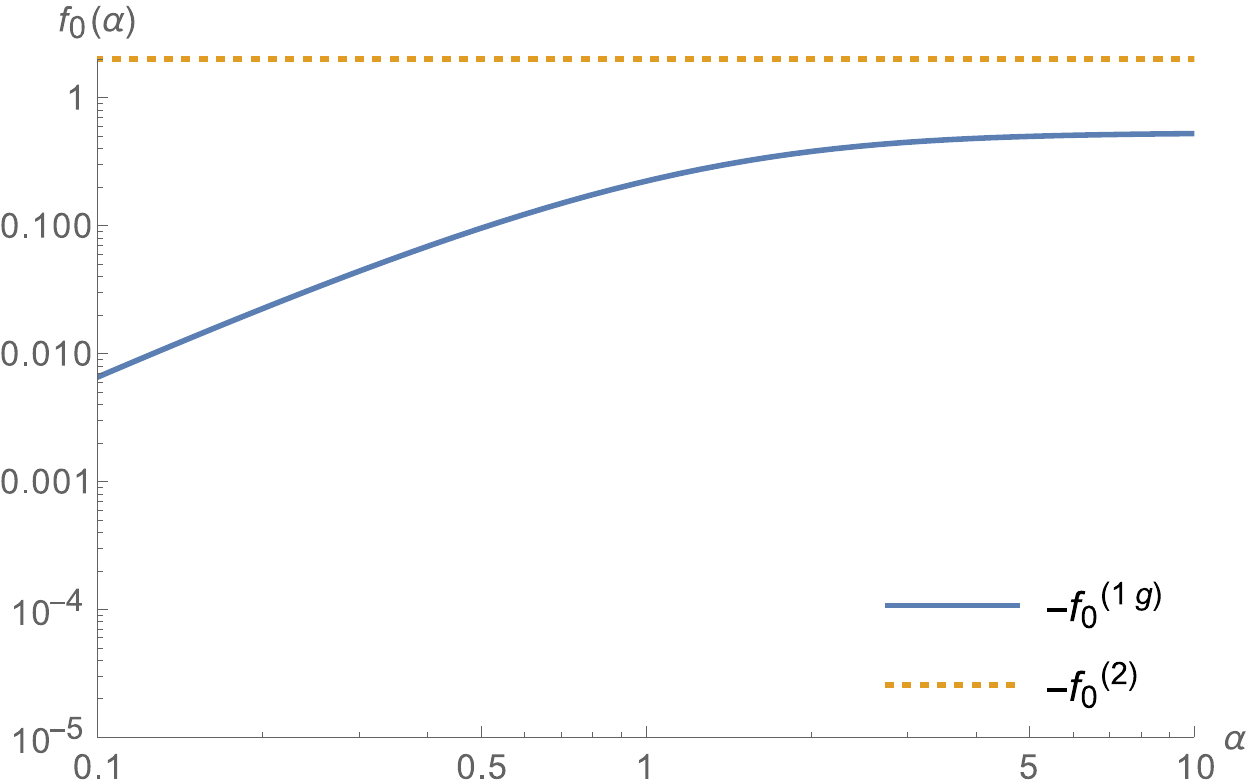}}
\\
\subfloat[spherical mass distribution, $n=1$]{\includegraphics[scale=.5]{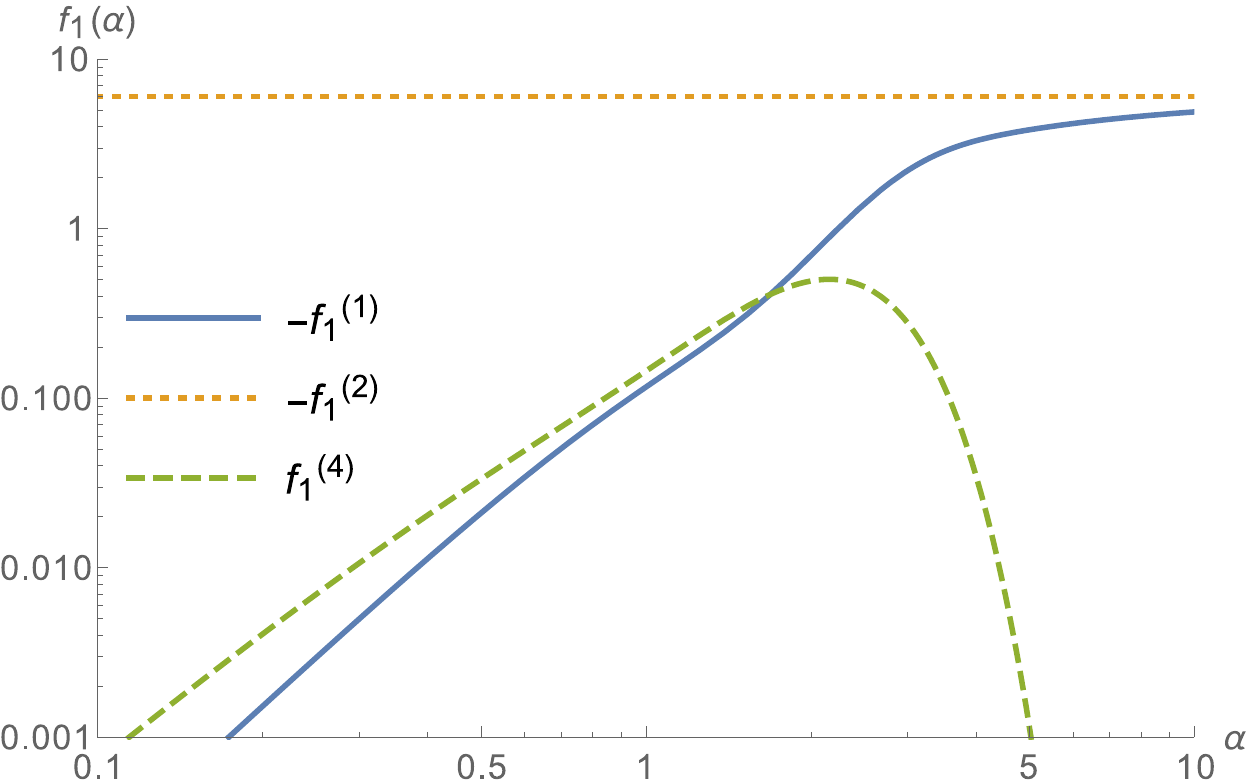}}
\hspace{1cm}
\subfloat[Gaussian mass distribution, $n=1$]{\includegraphics[scale=.5]{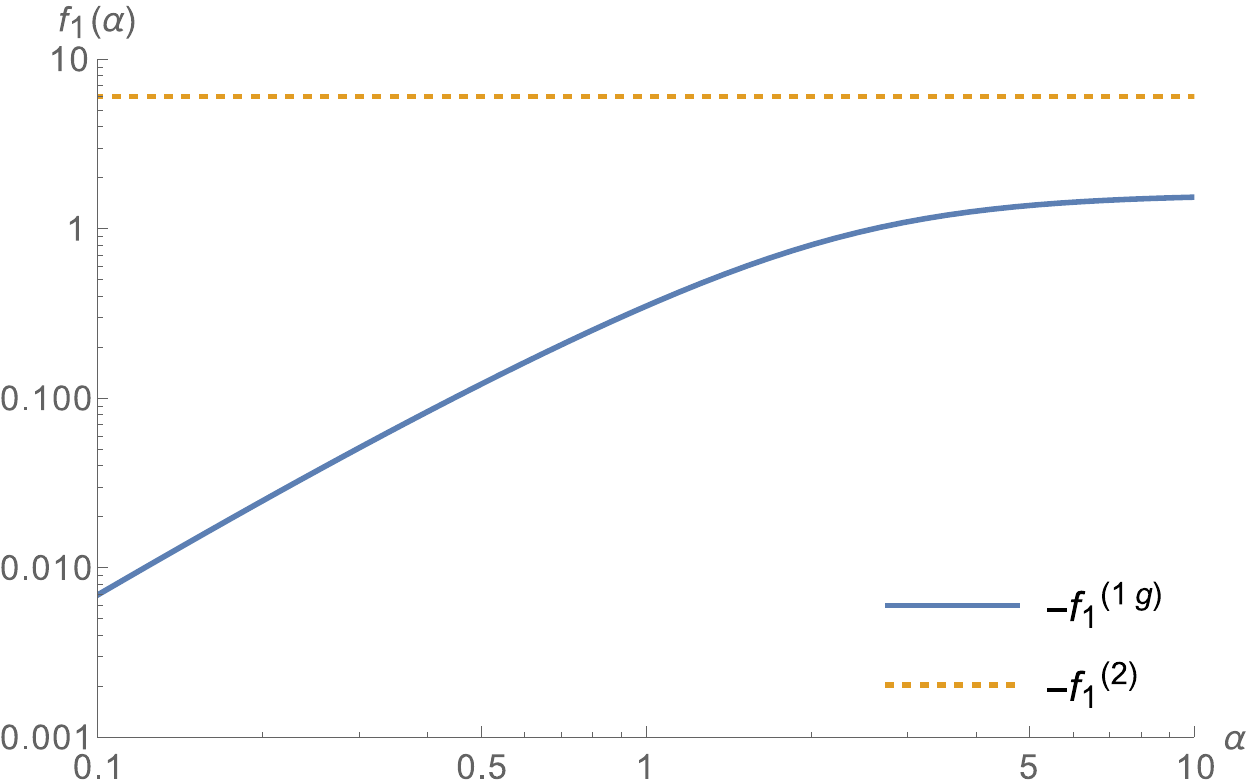}}
\caption{Comparison of the different terms contributing to $f_n(\alpha)$ for the spherical and Gaussian
atomic mass distribution, for the ground state as well as the first excited state.}
\label{fig:plot_f0_comparison}
\end{figure}

\subsection{Narrow wave-functions}
In the limit of large values for $\alpha$, i.\,e. narrow wave-functions, we can write
\begin{subequations}\begin{align}
f_n^\text{sphere}(\alpha) &\approx \sqrt{\frac{2}{\pi}}\,\int_0^\infty \D u\, \exp\left(-\frac{u^2}{2}\right) P_n(u) \,
\left( \frac{6}{5}\,\gamma_0\,\alpha^2 -2 \gamma_2 u^2 \right) \nnl
&= \frac{6}{5}\,\gamma_0\,\alpha^2 - 2\,\gamma_2\,(2n+1)
\intertext{for the spherical atomic mass distribution, and}
f_n^\text{Gauss}(\alpha) &\approx \sqrt{\frac{2}{\pi}}\,\int_0^\infty \D u\, \exp\left(-\frac{u^2}{2}\right) P_n(u) \,
\left( \frac{6}{5}\,\beta_0\,\alpha^2 -2 \beta_2 u^2 + \frac{\alpha^3}{2\,u}\,\erf\left(\sqrt{2}u\right) \right) \nnl
&= \frac{6}{5}\,\beta_0\,\alpha^2 - 2\,\beta_2\,(2n+1) + \sqrt{\frac{2}{\pi}}\,\left(\alpha^2 - \frac{2+4\,n}{3}\right)
\intertext{for the Gaussian distribution, and hence}
\tilde{f}^\text{sphere}_{n_1 n_2}(\alpha) &= 4\,(n_1-n_2)\,, \\
\tilde{f}^\text{Gauss}_{n_1 n_2}(\alpha) &= \frac{4}{3}\,\sqrt{\frac{2}{\pi}}\,(n_1-n_2)\,,
\end{align}\end{subequations}
where we used that $\gamma_2 \approx 1$ and $\beta_2 \approx 0$ for $R \gg a \gg \sigma$.
This yields exactly the previous results~\eqref{eqn:transition-energies-quadratic} with the
frequencies~\eqref{eqn:omega-sn-sphere} and~\eqref{eqn:omega-sn-gauss}.

\subsection{Intermediate wave-functions}\label{sec:spectrum-intermediate}
Now we want to go beyond the quadratic approximation for the potential, to see how the gravitational interaction
removes the degeneracy of the transition energies. For this, we consider the intermediate regime where
$\alpha$ is of the order of unity.
We are, again, interested in the case where $N \gg 1$ and $R \gg a \gg \sigma$. In this case, we have
$\gamma_0 \approx \beta_0$, $\gamma_i \approx 1 \, (i \geq 2)$, and $\varrho \gg 1$. We can then write for the
spherical atomic mass distribution
{\allowdisplaybreaks\begin{subequations}\begin{align}\label{eqn:fn-split-sphere}
f_n^\text{sphere}(\alpha) &\approx \frac{6}{5}\,\beta_0\,f_n^{(0)}(\alpha) + f_n^{(1)}(\alpha)
+ \frac{N\,\sigma^3}{R^3}\,f_n^{(2)}(\alpha)
+ \frac{N\,\sigma^3}{R^3}\,f_n^{(3)}(\alpha) + f_n^{(4)}(\alpha)
\intertext{with}
f_n^{(0)}(\alpha) &= \alpha^3\,\sqrt{\frac{2}{\pi}} \, \int_0^\infty \D \zeta \,
\exp \left(-\frac{\alpha^2\,\zeta^2}{2}\right) \, P_n(\alpha\,\zeta) = \alpha^2 \,, \\
f_n^{(1)}(\alpha) &= \alpha^3\,\sqrt{\frac{2}{\pi}} \, \int_0^1 \D \zeta \,
\exp \left(-\frac{\alpha^2\,\zeta^2}{2}\right) \, P_n(\alpha\,\zeta) \,
\left( -2\,\zeta^2 + \frac{3}{2}\,\zeta^3 -\frac{1}{5}\,\zeta^5 \right) \,, \\
f_n^{(2)}(\alpha) &= \lim_{\varrho \to \infty} \alpha^3\,\sqrt{\frac{2}{\pi}} \,\varrho^3\, \int_0^1 \D u \,
\exp \left(-\frac{\alpha^2\,\varrho^2\,u^2}{2}\right) \nnl
&\bleq \times P_n(\alpha\,\varrho\,u) \,
\left( -2\,u^2 + \frac{3}{2}\,u^3 -\frac{1}{5}\,u^5 \right) \,, \\
f_n^{(3)}(\alpha) &= \lim_{\varrho \to \infty} \alpha^3\,\sqrt{\frac{2}{\pi}} \,\varrho^3\, \int_1^\infty \D u \,
\exp \left(-\frac{\alpha^2\,\varrho^2\,u^2}{2}\right) \, P_n(\alpha\,\varrho\,u) \,\frac{1}{2\,u} = 0 \,, \\
f_n^{(4)}(\alpha) &= \lim_{\varrho \to \infty} \alpha^3\,\sqrt{\frac{2}{\pi}} \, \int_{1/\varrho}^1 \D u \,
\exp \left(-\frac{\alpha^2\,\varrho^2\,u^2}{2}\right) \, P_n(\alpha\,\varrho\,u) \,\frac{1}{2\,u} \,.
\end{align}\end{subequations}}%
These functions become $\varrho$-independent by taking only the zeroth order of the expansion around infinity.
$f_n^{(0)}$ is an $n$-independent term, which is large but does not contribute to the transition energies.
$f_n^{(3)}$ can be neglected since it goes to zero exponentially as $\varrho$ becomes large. The remaining terms,
$f_n^{(1)}$, $f_n^{(2)}$, and $f_n^{(4)}$, are of comparable size, cf. figure~\ref{fig:plot_f0_comparison}.
$f_n^{(2)}$ however enters into the full function $f_n$ with the small\footnote{This pre-factor is, e.\,g., less
than $10^{-4}$ for silicon and less than $10^{-5}$ for osmium at a few Kelvin.} pre-factor $N\,\sigma^3/R^3$,
and therefore can be neglected as well, allowing for a $\varrho$-independent approximation for $f_n$.
Figure~\ref{fig:plot_f0_comparison} shows that for $\alpha > 1$, $f_n^{(4)}$ is also negligible compared to
$f_n^{(1)}$. We nevertheless use both functions to calculate the transition energies with
\begin{equation}
\tilde{f}^\text{sphere}_{n_1 n_2}(\alpha)
= f_{n_2}^{(1)}(\alpha) + f_{n_2}^{(4)}(\alpha) - f_{n_1}^{(1)}(\alpha) - f_{n_1}^{(4)}(\alpha) \,.
\end{equation}
For the Gaussian atomic mass distribution one obtains instead
\begin{subequations}\begin{align}\label{eqn:fn-split-gauss}
f_n^\text{Gauss}(\alpha) &\approx \left(\frac{6}{5}\,\beta_0+\sqrt{\frac{2}{\pi}}\right)\,f_n^{(0)}(\alpha)
+ f_n^{(1g)}(\alpha)
+ \frac{N\,\sigma^3}{R^3}\,f_n^{(2)}(\alpha)
+ \frac{N\,\sigma^3}{R^3}\,f_n^{(3)}(\alpha)
\intertext{with}
f_n^{(1g)}(\alpha) &= \alpha^3\,\sqrt{\frac{2}{\pi}} \, \int_0^\infty \D \zeta \,
\exp \left(-\frac{\alpha^2\,\zeta^2}{2}\right) \, P_n(\alpha\,\zeta) \,
\left(\frac{\erf\left(\sqrt{2}\,\zeta\right)}{2\,\zeta} - \sqrt{\frac{2}{\pi}}\right) \,,
\end{align}\end{subequations}
were we accounted for the $n$-independent, $\sqrt{2/\pi}$-proportional, contribution of $f_n^{(1g)}$ in $f_n^{(0)}$. 
With the same arguments as before, this results in
\begin{equation}
\tilde{f}^\text{Gauss}_{n_1 n_2}(\alpha) = f_{n_2}^{(1g)}(\alpha) - f_{n_1}^{(1g)}(\alpha) \,.
\end{equation}
\begin{figure}
\centering
\subfloat[spherical atomic mass distribution]{\includegraphics[scale=.5]{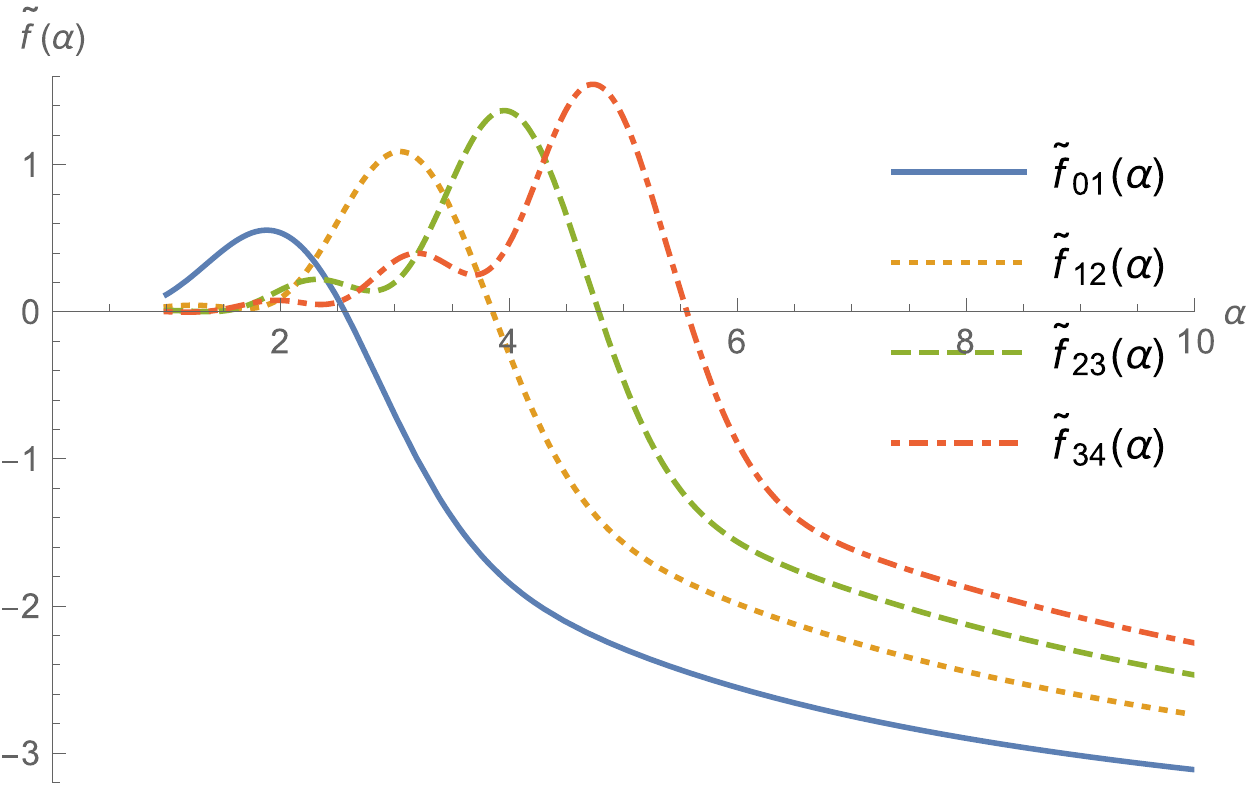}}
\hspace{1cm}
\subfloat[Gaussian atomic mass distribution]{\includegraphics[scale=.5]{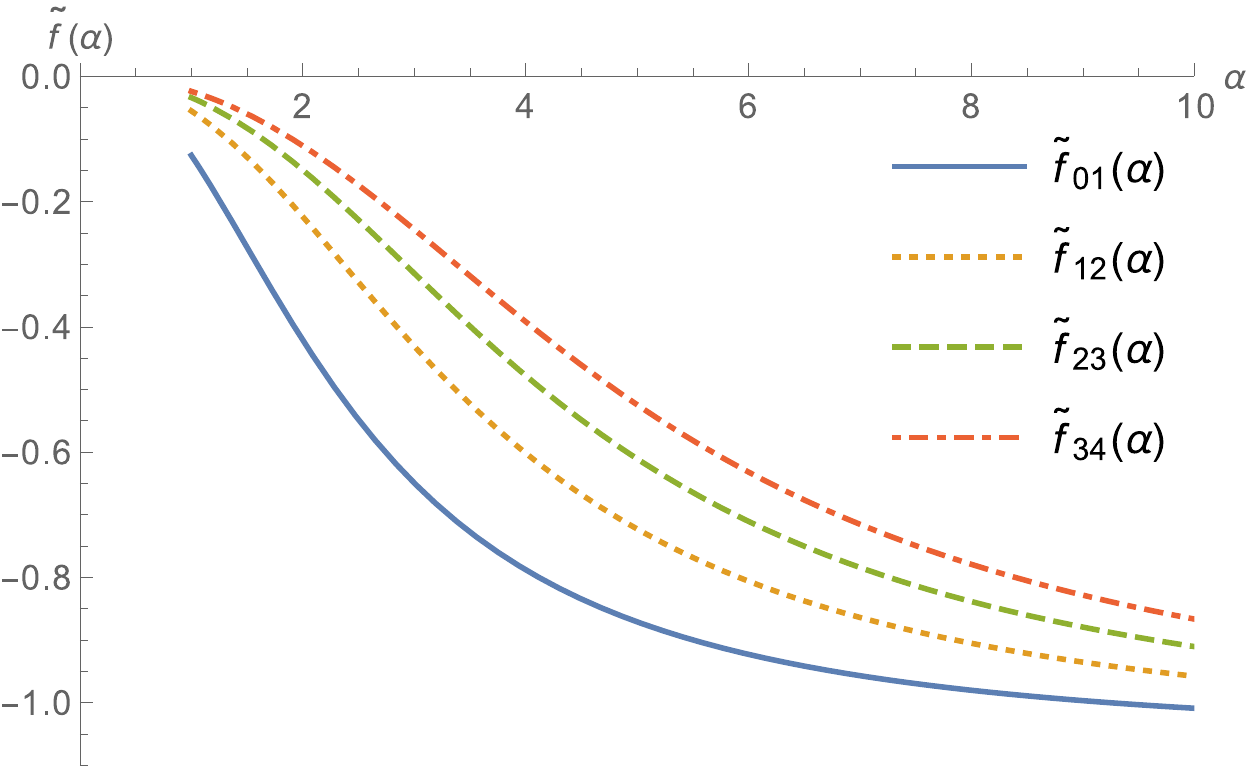}}
\caption{The coefficient function $\tilde{f}_{n_1 n_2}(\alpha)$ for the spherical and Gaussian
atomic mass distribution.}
\label{fig:plot_etrans}
\end{figure}%
In figure~\ref{fig:plot_etrans} these functions are plotted for the lowest four transitions with $\Delta n = 1$,
for both the spherical and Gaussian atomic mass distribution. One can see how the degeneracy is removed, and
there is a split of $\sim\order{1}$ for the spherical, and $\sim\order{0.1}$ for the Gaussian distribution,
respectively. In the limit of an infinitesimally narrow wave-function, i.\,e. $\alpha \to \infty$, all these
functions will converge against the same value, in agreement with equation~\eqref{eqn:transition-energies-quadratic}.

The order of the split of the spectral lines belonging to the same $\Delta n$ is given by the pre-factor in
equation~\eqref{eqn:transition-energies}. Taking, e.\,g., silicon at a few Kelvin (cf.~\cite{Yang:2013})
with $m_\text{atom} = \unit{28}{\atomicmass}$ and $\sigma \approx \unit{7.0 \times 10^{-12}}{\meter}$~\cite{Sears:1991},
we get
\begin{equation}\label{eqn:transition-energy-pre}
\frac{\Delta E^\text{trans}_{n, n+1}}{\hbar\,\omega_0} \sim \frac{G\,m_\text{atom}}{4\,\sigma^3\,\omega_0^2}
\approx \frac{0.0023}{\omega_0^2 / \second^{-2}}\,.
\end{equation}
The best value can be obtained for osmium with $m_\text{atom} = \unit{190}{\atomicmass}$ and
$\sigma \approx \unit{2.8 \times 10^{-12}}{\meter}$~\cite{Gao:1999}, where the above pre-factor is
two orders of magnitude larger.
\begin{figure}
\centering
\subfloat[pre-factor~\eqref{eqn:transition-energy-pre} of the split of the transition energies]{\includegraphics[scale=.5]{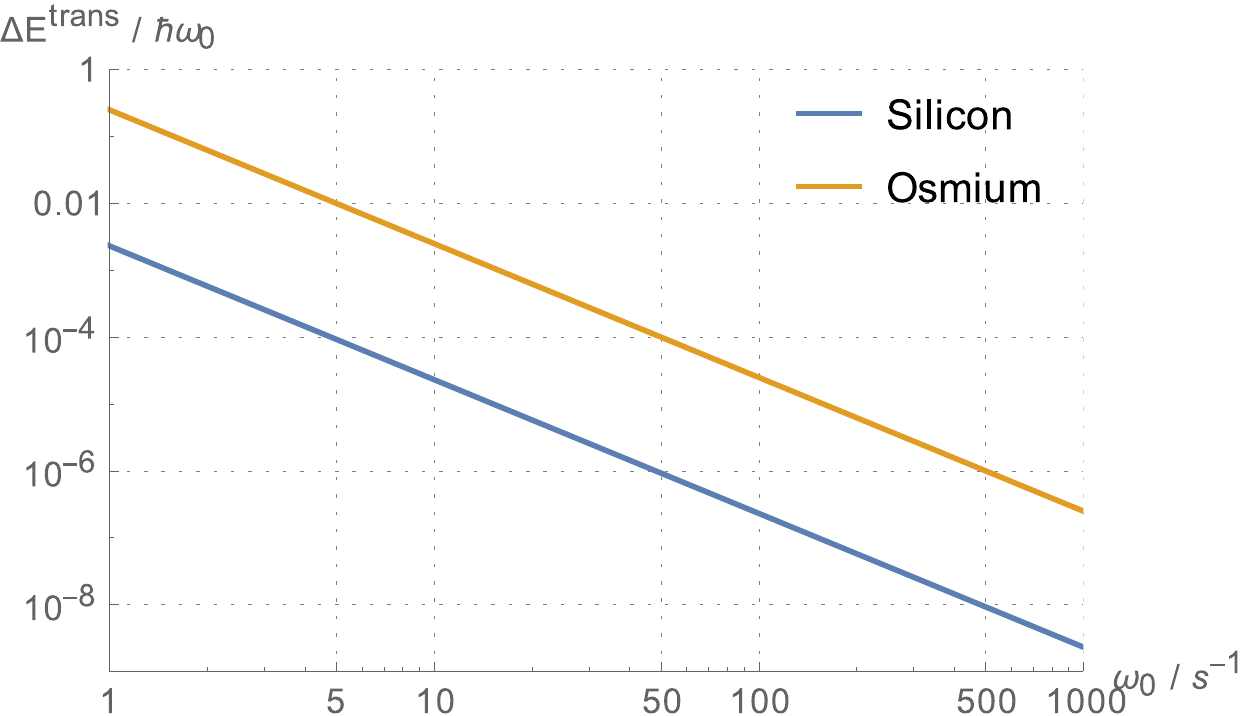}}
\hspace{1cm}
\subfloat[dependence between $m$ and $\omega_0$ for different values of $\alpha$]{\includegraphics[scale=.5]{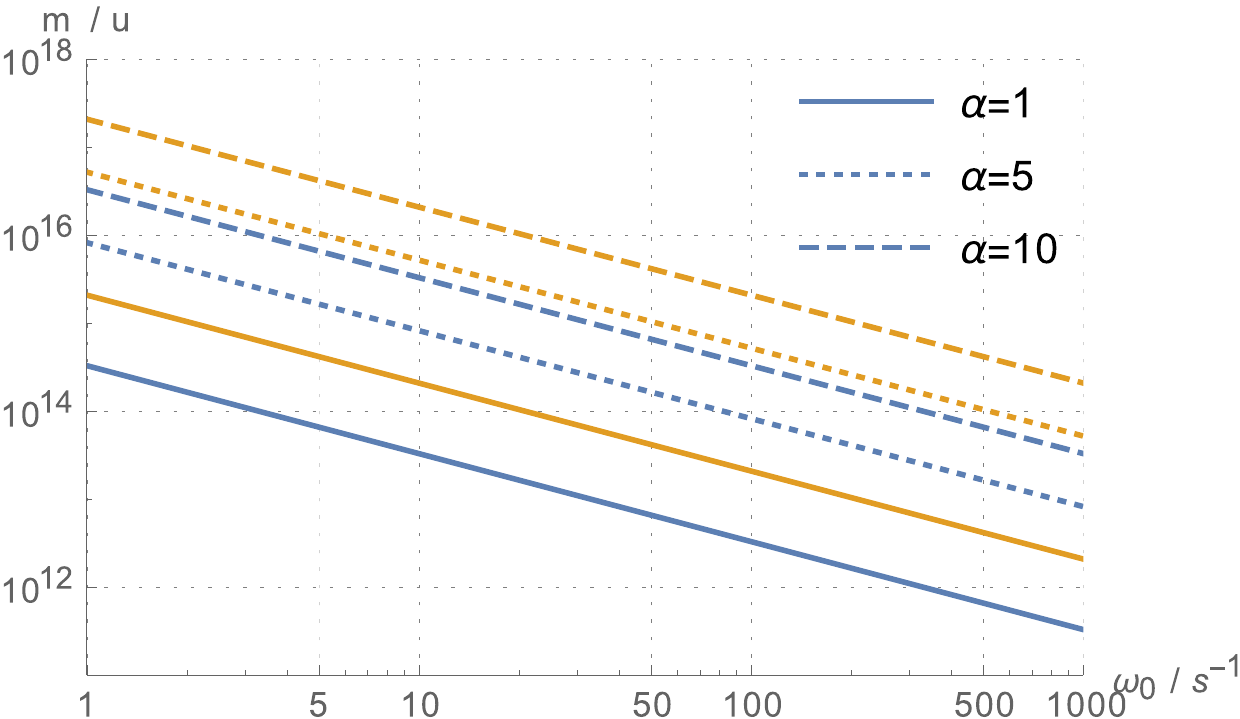}}
\caption{The first plot shows the dependence of the pre-factor~\eqref{eqn:transition-energy-pre}
of the split in transition energy on the trap frequency $\omega_0$. The second plot shows the
necessary mass for $\alpha$-values of 1, 5, and 10, respectively, for these frequencies $\omega_0$.
We used the values for silicon (in blue) and osmium (in orange) as given in the text.}
\label{fig:plot_etrans_m_alpha_omega}
\end{figure}%
Figure~\ref{fig:plot_etrans_m_alpha_omega} shows the dependence of this pre-factor on the trap frequency,
and the corresponding masses for different values of $\alpha$.

Qualitatively, the same effect can be expected from a three-dimensional harmonic oscillator, in situations where
the wave-function has comparable width in all directions, although the situation gets more complicated when
transitions are allowed in all three dimensions with different frequencies. We study the simpler case of an
axially symmetric state which is excited in only the longitudinal direction in appendix~\ref{app:axial}.

\subsection{Wide wave-functions}
\begin{figure}
\centering
\subfloat[pre-factor of the split of the transition energies for different frequencies]{\includegraphics[scale=.5]{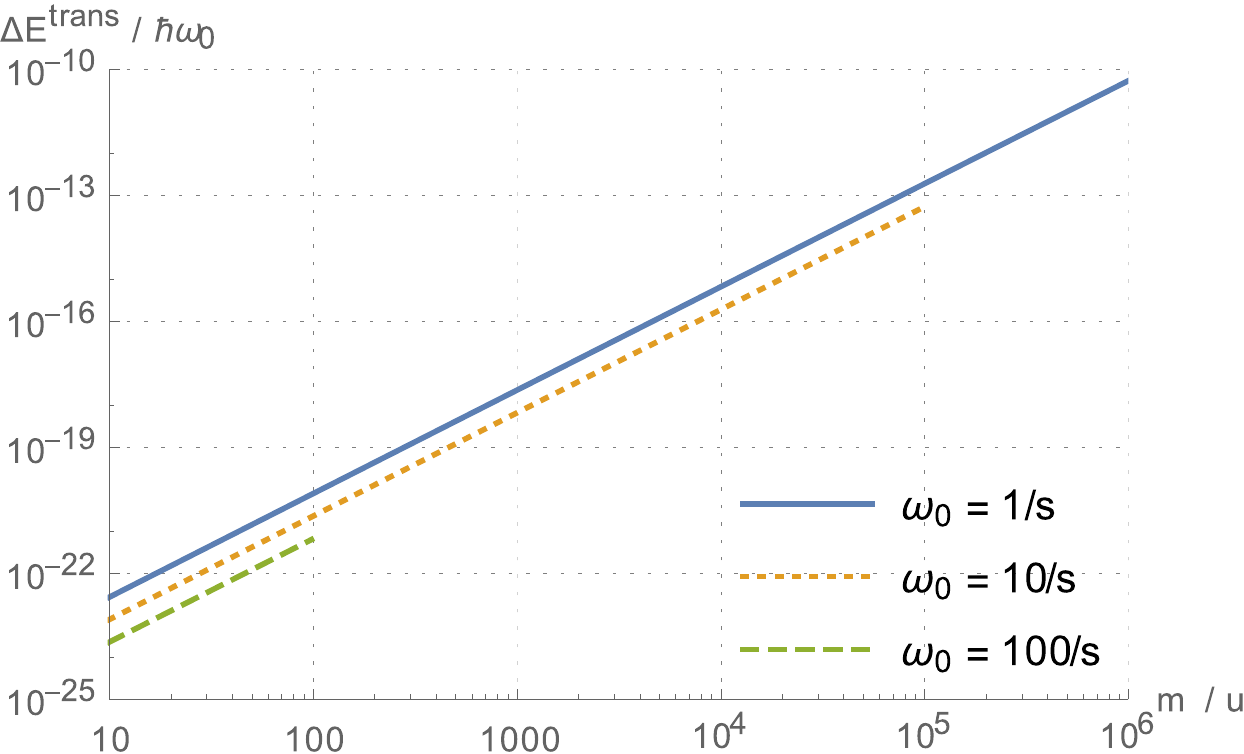}}
\hspace{1cm}
\subfloat[transition energies between different energy levels]{\includegraphics[scale=.5]{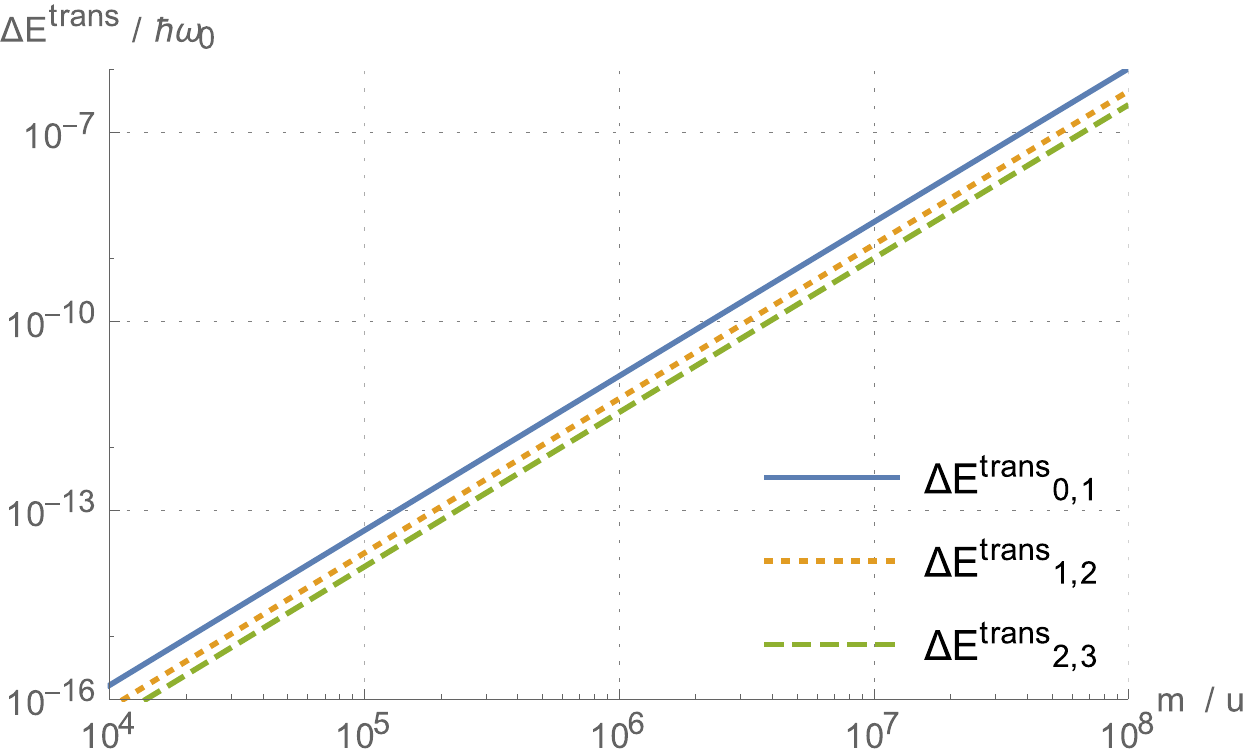}}
\caption{Dependence of the transition energies for wide wave-functions on the mass.
The first plot shows the mass dependence of the pre-factor in equation~\eqref{eqn:transition-energies-wide} for
different trap frequencies. The second plot shows the mass dependence for the actual transition energies between the
first four energy levels for a trap frequency $\omega_0 = \unit{1}{\second^{-1}}$.
We used $\sigma \approx \unit{7.0 \times 10^{-12}}{\meter}$ for silicon, as in the text.
The plotted lines end for the mass value for which the wave-function width equals the radius $R$ of the microsphere.}
\label{fig:plot_etrans_wide}
\end{figure}

In the limit of a wide wave-function, $\alpha \to 0$ and $m\,\omega_0\,R^2 \ll \hbar$, the dominant contribution
comes from the function $f_n^{(3)}$, yielding
\begin{subequations}\begin{align}\label{eqn:energy-correction-wide}
f_n(\alpha) &\approx -\frac{\hbar}{4\,\sqrt{2\,\pi}\,m_\text{atom}\,\sigma^2\,\omega_0}\,F_n \, \alpha^5 \,\ln\alpha
\intertext{with the $n$-dependent pre-factors}
F_n &= -\lim_{\alpha \to 0} \frac{1}{\ln\alpha} \,
\int_0^\infty \D u\, \frac{\exp\left(-\frac{u^2}{2}\right)}{u}\, P_n(u) \,.
\end{align}\end{subequations}
The first six values for the pre-factors are
\begin{equation}
F_0 = 1\,, \quad
F_1 = \frac{3}{4}\,, \quad
F_2 = \frac{41}{64}\,, \quad
F_3 = \frac{147}{256}\,, \quad
F_4 = \frac{8649}{16384}\,, \quad
F_5 = \frac{32307}{65536}\,.
\end{equation}
The resulting transition energies are
\begin{equation}\label{eqn:transition-energies-wide}
E^\text{trans}_{n_1 n_2} = \hbar\,\omega_0\, \left[n_2 - n_1
+ G\,\sqrt{\frac{2\,m^5}{\pi\,\hbar^3\,\omega_0}}\,\ln\left(2\,\sigma\,\sqrt{\frac{m\,\omega_0}{\hbar}}\right)\,
\left(F_{n_2}-F_{n_1}\right) \right]\,.
\end{equation}
$F_{n_2}-F_{n_1}$ is of the order of unity. The magnitude of the pre-factor for different trap frequencies, as
well as the transition energies for a fixed frequency of $\unit{1}{\second^{-1}}$ are plotted in
figure~\ref{fig:plot_etrans_wide}. Since for the same trap frequencies the mass must be smaller in order to still
be in the wide wave-function regime, the transition energies are several orders of magnitude below those for
the intermediate regime.

There is also a ``semi-wide'' regime where the wave-function width is between $\sigma$ and $R$.
We omit the detailed discussion of this regime here. This has no effect on the qualitative results provided,
although this regime could in principle be treated along the same lines.
Experimentally, the narrow and intermediate regime are the most relevant for trapped microspheres.
We mainly discussed the wide wave-function here for reasons of completeness, and because this is the
situation at hand in experimental tests based on molecular interferometry.

\section{Gravitational dynamics of squeezed coherent Gaussian states}\label{sec:dynamics}
Inspired by the proposal by Yang et~al.~\cite{Yang:2013}, here we discuss the dynamical properties of a trapped
microsphere that has been prepared in a squeezed ground state.
A particular property of a harmonic potential is that a Gaussian wave packet remains Gaussian during its
time evolution. This is because a Gaussian is fully determined by the first and second moments,
$\langle x \rangle$, $\langle p \rangle$, $\langle x^2 \rangle$, $\langle p^2 \rangle$, and the correlation
$\langle x\,p+p\,x \rangle$, and the \schr\ equation gives a closed system of equations for the same.

This property does not persist if a non-quadratic potential, such as our gravitational potential $V_g$, is added
to the Hamiltonian. However, since the potential $V_g$ is usually weak compared to the harmonic trap potential,
we can assume that the dynamics of an initially Gaussian wave packet are still approximately determined by
the time evolution of the first and second moments~\cite{Yang:2013,Giulini:2014,Colin:2014}.

With the general \schr\ equation~\eqref{eqn:sn-1d} one gets for the first moments~\cite{Giulini:2014}
\begin{subequations}\label{eqn:system-first-moments}\begin{align}
\frac{\partial}{\partial t}\,\langle x \rangle &= \frac{1}{m}\,\langle p \rangle \\
\frac{\partial}{\partial t}\,\langle p \rangle &= -m\,\omega_0^2\,\langle x \rangle
- \left\langle \frac{\partial V_g}{\partial x} \right\rangle \,.
\end{align}\end{subequations}
Since $I_{\rho_c}$ is an even function, and hence its derivative is odd, the expectation value of the derivative
of $V_g$ vanishes for any wave-function and any mass density function $\rho_c$. Therefore, the time evolution
of the first moments remains completely unchanged by self-gravitation, as one would intuitively expect,
and in agreement with the Ehrenfest theorem.

For the second moments, first define the three-dimensional vector $\widetilde{\vec u}$ with components
\begin{subequations}\begin{align}
\widetilde{u}_1(t) &= \langle x^2 \rangle - \langle x \rangle^2 \\
\widetilde{u}_2(t) &= \frac{1}{m^2}\,\left( \langle p^2 \rangle - \langle p \rangle^2 \right) \\
\widetilde{u}_3(t) &= \frac{1}{m}\,\left( \langle x\,p+p\,x \rangle - 2\,\langle x \rangle\,\langle p \rangle \right) \,. 
\end{align}\end{subequations}
Then the \schr\ equation~\eqref{eqn:sn-1d} yields the following system of equations:
\begin{subequations}\begin{align}\label{eqn:sn-system-1}
\frac{\D}{\D t}\,\widetilde{\vec u}(t) &=
\begin{pmatrix}
0 & 0 & 1 \\
0 & 0 & -\omega_0^2 \\
-2\,\omega_0^2 & 2 & 0
\end{pmatrix}\,\widetilde{\vec u}(t)
+ \begin{pmatrix}
0 \\
h(t) \\
\widetilde{g}(t)
\end{pmatrix}
\intertext{with}
\widetilde{g}(t) &= -\frac{2}{m}\,\left\langle x \, \frac{\partial V_g}{\partial x} \right\rangle \\
h(t) &= -\frac{1}{m^2}\,\left\langle p \, \frac{\partial V_g}{\partial x}
+ \frac{\partial V_g}{\partial x}\,p \right\rangle \label{eqn:function-h} \,.
\end{align}\end{subequations} 
The function $h(t)$ can be shown to equal (see appendix~\ref{app:function-h})
\begin{equation}
h(t) = -\frac{1}{m}\,\frac{\partial}{\partial t}\langle V_g \rangle \,.
\end{equation}
Redefining $u_1 = \widetilde{u_1}$, $u_2 = \widetilde{u}_2 + \frac{1}{m}\,\langle V_g \rangle$,
$u_3 = \widetilde{u_3}$, and
\begin{equation}\label{eqn:function-g}
g(t) = \widetilde{g}(t) - \frac{2}{m}\,\langle V_g \rangle
= -\frac{2}{m} \,\left(\left\langle x \, \frac{\partial V_g}{\partial x} \right\rangle
+ \langle V_g \rangle\right) \,,
\end{equation}
the system~\eqref{eqn:sn-system-1} then takes the form
\begin{equation}\label{eqn:sn-system-2}
\frac{\D}{\D t}\,\vec u(t) =
\begin{pmatrix}
0 & 0 & 1 \\
0 & 0 & -\omega_0^2 \\
-2\,\omega_0^2 & 2 & 0
\end{pmatrix}\,\vec u(t)
+ \begin{pmatrix}
0 \\
0 \\
g(t)
\end{pmatrix} \,,
\end{equation}
and is equivalent (given corresponding initial conditions) to the third order equation for $u_1$
\begin{equation}\label{eqn:third-order-ode}
\frac{\D^3 u_1(t)}{\D t^3} + 4\,\omega_0^2\,\frac{\D u_1(t)}{\D t}
= \frac{\partial g(t)}{\partial t}\,. 
\end{equation}
Up to this point, no restrictions have been imposed on the shape of the wave-function.
Note, however, that $g(t)$ depends on the wave-function. Hence, the right-hand-side is not a mere inhomogeneity
but it renders the equation nonlinear, and in general the system~\eqref{eqn:sn-system-2} will not be closed. 
If now we assume that the wave-function is of Gaussian shape,
\begin{equation}\label{eqn:gaussian-wf}
\psi(t,x) = \left(2\pi\,u_1(t)\right)^{-1/4}\,\exp\left(-\frac{(x-\langle x \rangle(t))^2}{4\,u_1(t)}
+\frac{\rmi}{\hbar}\,\langle p \rangle(t) \, x + \rmi\,\varphi\right)\,,
\end{equation}
then the gravitational potential $V_g$ is completely determined\footnote{for a given solution of the 
system~\eqref{eqn:system-first-moments} for the first moments} by the wave-function width $u_1(t)$, and so is the
function $g(t)$. We then obtain instead of equation~\eqref{eqn:third-order-ode} the closed equation
\begin{equation}\label{eqn:third-order-ode-2}
\frac{\D^3 u_1(t)}{\D t^3} + \left( 4\,\omega_0^2
- g'(u_1(t))\right)\,\frac{\D u_1(t)}{\D t} = 0\,,
\end{equation}
where the prime denotes the derivative by $u_1$. In this case, the gravitational interaction acts like a
wave-function width dependent change of the frequency of the internal oscillations of a Gaussian state.

Obviously, such internal oscillations appear only in the case of a squeezed state---for a coherent ground state,
$u_1$ would be a constant in time. Without the gravitational interaction, starting initially with a state whose width
is $\kappa$ times the width of the ground state, the solution of equation~\eqref{eqn:third-order-ode-2} is
\begin{equation}
u_1(t) = \frac{\hbar}{2\,m\,\omega_0}\,\left(\kappa^2\,\cos^2\omega_0\,t
+ \frac{1}{\kappa^2}\,\sin^2\omega_0\,t\right)\,.
\end{equation}
Hence, without gravity the width of the wave-function and $\langle x \rangle$ oscillate in phase. The gravitational
interaction only affects the oscillation frequency of $u_1$, and not that of $\langle x \rangle$, and therefore
induces a de-phasing. This can, in principle, be observed experimentally.

In order to obtain quantitative results, we calculate the function $g$. Inserting the gravitational
potential~\eqref{eqn:axial-grav-potential} and the wave-function~\eqref{eqn:gaussian-wf} into~\eqref{eqn:function-g}
yields
\begin{align}
g(u_1(t)) &= \frac{G}{\pi\,m\,u_1}\,\int_{-\infty}^\infty \D x\,\exp\left(-\frac{(x-\langle x \rangle)^2}{u_1}\right)
\int_0^\infty \D d\,\exp\left(-\frac{d^2}{2\,u_1}\right)\nnl
&\bleq \times \Bigg[x\,I_{\rho_c}'(d) \left(\exp\left(\frac{d\,(x-\langle x \rangle)}{u_1}\right)
-\exp\left(\frac{d\,(x-\langle x \rangle)}{u_1}\right)\right)\nnl
&\bleq +I_{\rho_c}(d) \left(\exp\left(\frac{d\,(x-\langle x \rangle)}{u_1}\right)
+\exp\left(\frac{d\,(x-\langle x \rangle)}{u_1}\right)\right)\Bigg]\,.
\end{align}
Introducing dimensionless variables, as in the previous section,
\begin{equation}
\xi = \frac{x-\langle x \rangle}{\sqrt{u_1}} \,, \quad\quad
\zeta = \frac{d}{2\,\sigma} \,, \quad\quad
\alpha = \frac{2\,\sigma}{\sqrt{u_1}} \,, \quad\quad
\varrho = \frac{R}{\sigma} = \sqrt[3]{\frac{3\,m}{4\pi\,\rho\,\sigma^3}} \,,
\end{equation}
this can be rewritten as
\begin{align}
g(\alpha) &= \frac{G\,m_\text{atom}\,\alpha}{\pi\,\sigma}\,\int_{-\infty}^\infty \D \xi\,\rme^{-\xi^2}
\int_0^\infty \D \zeta\,\rme^{-\frac{\alpha^2\,\zeta^2}{2}}\nnl
&\bleq \times \Bigg[\left(\frac{\xi}{\alpha}+\frac{\langle x \rangle}{2\,\sigma}\right)\,i'(\zeta,\varrho)
\left(\rme^{\alpha\,\zeta\,\xi}-\rme^{-\alpha\,\zeta\,\xi}\right)
+i(\zeta,\varrho) \left(\rme^{\alpha\,\zeta\,\xi}+\rme^{-\alpha\,\zeta\,\xi}\right)\Bigg]\,,
\end{align}
where $i(\zeta,\varrho)$ is defined as in the previous section, and $i'(\zeta,\varrho)$ denotes the derivative
by $\zeta$. Evaluating the $\xi$-integral and taking the derivative leads to the desired function
\begin{align}\label{eqn:g-prime}
g'(u_1(t)) &= -\frac{\alpha^3}{8\,\sigma^2}\,\frac{\partial g(\alpha)}{\partial \alpha}\nnl
&= -\frac{G\,m_\text{atom}\,\alpha^3}{16\,\sqrt{\pi}\,\sigma^3}\,
\int_0^\infty \D\zeta\, \rme^{-\frac{\alpha^2\zeta^2}{4}} \,
\left(\zeta\,i'(\zeta,\varrho) + 2\,i(\zeta,\varrho)\right)\,\left(2-\alpha^2\,\zeta^2\right)\,.
\end{align}
As before, we discuss the limits of a narrow and wide wave-function, and the intermediate regime.

\subsection{Narrow wave-functions}
First we consider the limit $\alpha \to \infty$, corresponding to a narrow wave-function. We have
\begin{subequations}\begin{align}
\left[ \zeta\,i'(\zeta,\varrho) + 2\,i(\zeta,\varrho) \right]^\text{sphere} &= 
\begin{cases}
\frac{12}{5} \gamma_0 - 8 \gamma_2 \zeta^2 +\frac{15}{2} \gamma_3 \zeta^3-\frac{7}{5} \gamma_5 \zeta^5
&\text{for}\ \zeta\leq 1\,,\\
\frac{1}{2\,\zeta} + \frac{12}{5} \beta_0 - 8 \beta_2 \zeta^2 +\frac{15}{2} \beta_3 \zeta^3
-\frac{7}{5} \beta_5 \zeta^5 &\text{for}\ \zeta\leq \varrho\,,\\
\frac{N+1}{2\,\zeta} &\text{for}\ \zeta \geq \varrho
\end{cases}
\intertext{and}
\left[ \zeta\,i'(\zeta,\varrho) + 2\,i(\zeta,\varrho) \right]^\text{Gauss} &=
\sqrt{\frac{2}{\pi}}\,\rme^{-2\,\zeta^2} + \frac{\erf\left(\sqrt{2}\,\zeta\right)}{2\,\zeta} \nnl
&\bleq + \begin{cases}
\frac{12}{5} \beta_0 - 8 \beta_2 \zeta^2 +\frac{15}{2} \beta_3 \zeta^3
-\frac{7}{5} \beta_5 \zeta^5 &\text{for}\ \zeta\leq \varrho\,,\\
\frac{N}{2\,\zeta} &\text{for}\ \zeta \geq \varrho
\end{cases}\\
&= 2\,\sqrt{\frac{2}{\pi}} + \frac{12}{5} \beta_0
- 8\,\left(\frac{1}{3}\,\sqrt{\frac{2}{\pi}}+\beta_2\right)\,\zeta^2 + \order{\zeta^3} \,.
\end{align}\end{subequations}
As before, we use that $\gamma_0 \approx \beta_0$, $\gamma_i \approx 1 \, (i \geq 2)$, and $\beta_2 \ll 1$
for $N \gg 1$ and $\varrho \gg 1$. With this, to lowest order one simply obtains
\begin{equation}\label{eqn:relation-g-omega-sn}
g'(u_1(t)) = -4\,\omega_\text{SN}^2\,
\end{equation}
with the respective values~\eqref{eqn:omega-sn-sphere} and~\eqref{eqn:omega-sn-gauss} for the spherical and the
Gaussian mass distributions.

Hence, we recover the result from~\cite{Yang:2013}, that for a narrow wave-function the \schr--Newton interaction
yields a frequency shift to
\begin{equation}
\omega = \sqrt{\omega_0^2 + \omega_\text{SN}^2}
\end{equation}
for the internal oscillations.

\subsection{Intermediate wave-functions}\label{sec:dynamics-inter}
\begin{figure}
\centering
\subfloat[Narrow wave-functions]{\includegraphics[scale=.5]{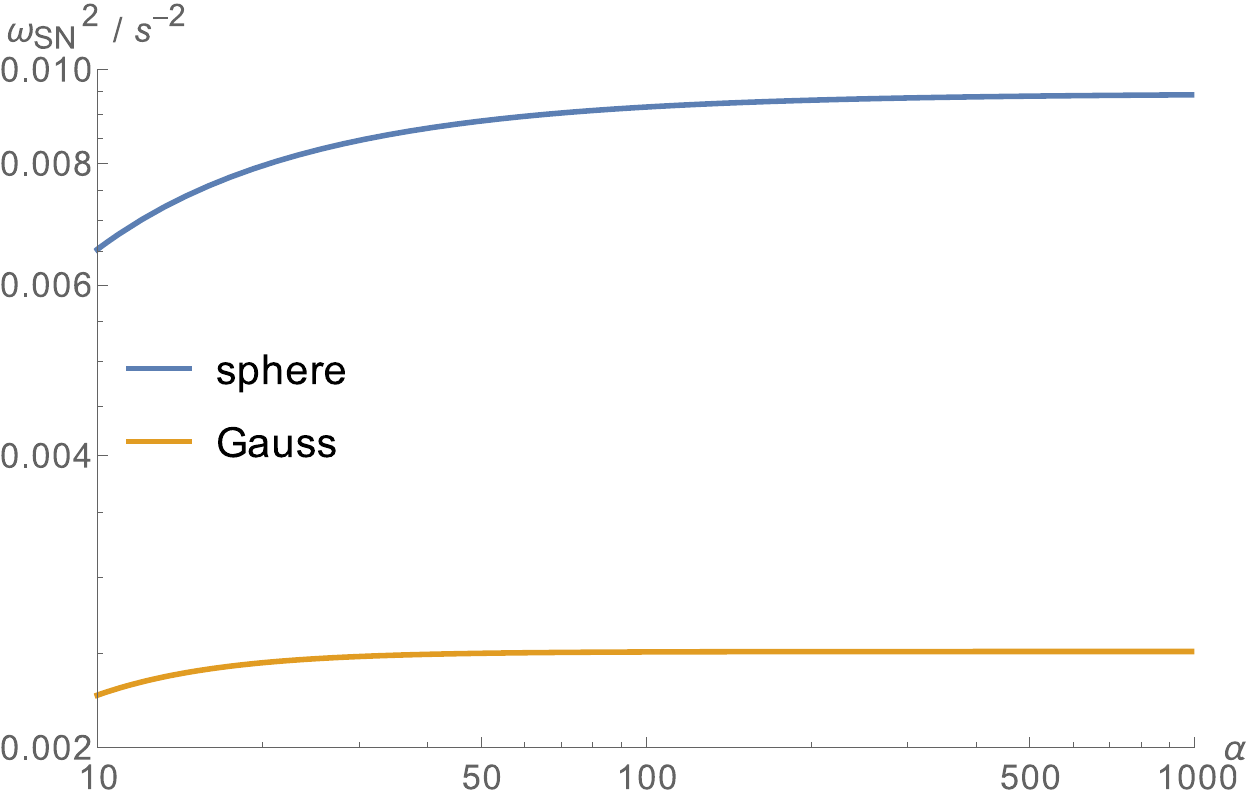}}
\hspace{1cm}
\subfloat[Intermediate wave-functions]{\includegraphics[scale=.5]{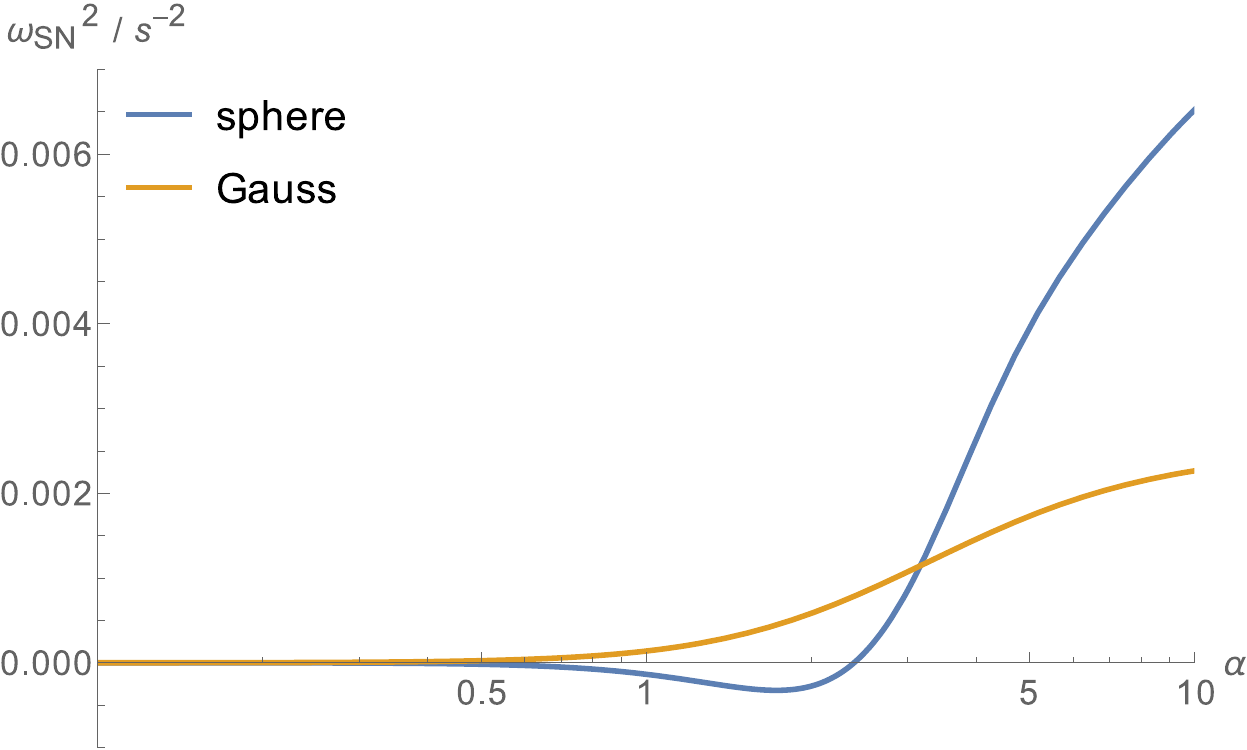}}
\caption{The plots show the dependence of the gravitational correction of the internal oscillation frequency
in the regimes of narrow and intermediate wave-functions, respectively, with respect to the parameter $\alpha$.
Values are for silicon at 10\,\kelvin, as in reference~\cite{Yang:2013}.}
\label{fig:plot_dynamics}
\end{figure}
If $\alpha$ approaches values of the order of unity, we can split up the integral in a similar way as in
subsection~\ref{sec:spectrum-intermediate}. Making use of $\gamma_0 \approx \beta_0$, one obtains
{\allowdisplaybreaks\begin{subequations}\begin{align}
g'_\text{sphere}(u_1(t)) &= -\frac{G\,m_\text{atom}}{16\,\sqrt{\pi}\,\sigma^3}\,
\left(\frac{12}{5}\,\beta_0\,k^{(0)} + k^{(1)} + \frac{N\,\sigma^3}{R^3}\,k^{(2)}
 + \frac{N\,\sigma^3}{R^3}\,k^{(3)}  + k^{(4)}\right)
\intertext{and}
g'_\text{Gauss}(u_1(t)) &= -\frac{G\,m_\text{atom}}{16\,\sqrt{\pi}\,\sigma^3}\,
\left(\frac{12}{5}\,\beta_0\,k^{(0)} + k^{(1,g)} + \frac{N\,\sigma^3}{R^3}\,k^{(2)}
 + \frac{N\,\sigma^3}{R^3}\,k^{(3)}\right)
\intertext{with}
k^{(0)} &= \alpha^3\,\int_0^\varrho \D\zeta\,\exp\left(-\frac{\alpha^2\zeta^2}{4}\right)\,\left(2-\alpha^2\zeta^2\right)
= 2\,\alpha^3\,\varrho\,\exp\left(-\frac{\alpha^2\varrho^2}{4}\right) \\
k^{(1)} &= \alpha^3\,\int_0^1 \D\zeta\,\exp\left(-\frac{\alpha^2\zeta^2}{4}\right)\,\left(2-\alpha^2\zeta^2\right)\,
\left(-8\,\zeta^2 + \frac{15}{2}\,\zeta^3 - \frac{7}{5}\,\zeta^5\right) \\
k^{(1,g)} &= \alpha^3\,\int_0^\infty \D\zeta\,\exp\left(-\frac{\alpha^2\zeta^2}{4}\right)\,
\left(2-\alpha^2\zeta^2\right)\,\left(\sqrt{\frac{2}{\pi}}\,\rme^{-2\,\zeta^2}
+ \frac{\erf\left(\sqrt{2}\,\zeta\right)}{2\,\zeta}\right) \\
k^{(2)} &= \alpha^3\,\int_0^\varrho \D\zeta\,\exp\left(-\frac{\alpha^2\zeta^2}{4}\right)
\left(2-\alpha^2\zeta^2\right)
\left(-\frac{8\zeta^2}{\varrho^3} + \frac{15\zeta^3}{2\varrho^4} - \frac{7\zeta^5}{5\varrho^6}\right) \\
k^{(3)} &= \alpha^3\,\int_\varrho^\infty \D\zeta\,\exp\left(-\frac{\alpha^2\zeta^2}{4}\right)\,
\left(2-\alpha^2\zeta^2\right)\,\frac{1}{2\,\zeta} \\
k^{(4)} &= \alpha^3\,\int_1^\infty \D\zeta\,\exp\left(-\frac{\alpha^2\zeta^2}{4}\right)\,
\left(2-\alpha^2\zeta^2\right)\,\frac{1}{2\,\zeta} \,.
\end{align}\end{subequations}}%
For $\varrho \gg 1$ both $k^{(0)}$ and $k^{(3)}$ can be neglected. For $\varrho \to \infty$,
$k^{(2)} \to 64\,\sqrt{\pi}$, but since $k^{(2)}$ is multiplied with the small pre-factor $N\,\sigma^3/R^3$
it can be neglected in comparison to $k^{(1)}$, $k^{(1,g)}$, and $k^{(4)}$ as well, just like $f_n^{(2)}$ in
subsection~\ref{sec:spectrum-intermediate}. The resulting $\omega_\text{SN}^2$, according to
equation~\eqref{eqn:relation-g-omega-sn}, is plotted as a function of the parameter $\alpha$ in
figure~\ref{fig:plot_dynamics}, for both the spherical and the Gaussian mass distribution, in the regimes
of narrow ($\alpha \gg 1$) and intermediate wave-functions ($\alpha \approx 1$).
One can see from figure~\ref{fig:plot_dynamics}a that the values~\eqref{eqn:omega-sn-sphere}
and~\eqref{eqn:omega-sn-gauss} are recovered in the limit $\alpha \to \infty$.

The $\alpha$-dependence of $\omega_\text{SN}^2$ turns equation~\eqref{eqn:third-order-ode-2} into a nonlinear
differential equation for the wave-function width. However, for finite values of $\alpha$, $\omega_\text{SN}^2$
only becomes smaller compared to the narrow wave-function case. Therefore, in order to experimentally
observe the frequency shift for the internal oscillations, the wave-function should be as narrow as possible,
contrary to the energy spectrum, where we found the most significant effect in the intermediate regime.

\subsection{Wide wave-functions}
Finally, in the limit $\alpha \to 0$, i.\,e. for very wide wave-functions, the dominant contribution comes from
$k^{(3)}$, yielding
\begin{equation}
 g'(u_1(t)) \xrightarrow{\alpha \to 0} \frac{G\,m_\text{atom}}{16\,\sqrt{\pi}\,\sigma^3}\,
\varrho^3\,\alpha^3\,\ln\alpha
\approx -\frac{3\,G\,m_\text{atom}\,m}{16\,\sqrt{\pi^3}\,\rho\,\sigma^3}\,u_1^{-3/2}\,\ln\frac{u_1}{\sigma^2} \,.
\end{equation}
Inserting this result into~\eqref{eqn:third-order-ode-2} yields a nonlinear differential equation, whose solution
gives the deviation from the behaviour without gravity. The effect is, however, much smaller for the wide
wave-function than in the case of narrow wave-functions.

\section{Conclusions}
In this paper we provided a thorough survey of the effects of the gravitational self-interaction, described by the \sne,
on both the stationary states and the dynamics of a micron-sized sphere in a harmonic trap potential. We took the
finite size of the system into account, as well as its crystalline substructure, and discussed the results for the
different regimes of a wave-function that is wide, narrow, and comparable in width with the localisation of the nuclei
in the crystal.

For the dynamics of a squeezed Gaussian state we recover the result from~\cite{Yang:2013}, that for a narrow state
there is a frequency shift for the internal oscillations, and hence a de-phasing compared to the oscillations of the
centre, $\langle x \rangle$, of the wave-function. The conclusion by Yang et\,al.~\cite{Yang:2013} was that for a
silicon crystal at 10\,\kelvin\ and a trap frequency of $2\,\pi \times 10\,\second^{-1}$ a quality factor of
$Q \gtrsim 3 \times 10^6$ would be required for an experimental test of the \schr--Newton effect.

Here, we could show that this result in the limit of a narrow wave-function is a best case scenario, in the sense that
for a wider wave-function the de-phasing between internal and external oscillations only becomes smaller.
We conclude from our considerations in section~\ref{sec:dynamics-inter} that for the given values in
reference~\cite{Yang:2013} a minimum mass of about $10^{15}$ atomic mass units is required.
Below that mass, i.\,e. for $\alpha \lesssim 10$, the amount of de-phasing drops significantly.

Contrary to this, we found in section~\ref{sec:spectrum} that the \schr--Newton effect on the energy spectrum 
is most pronounced in the regime of intermediate wave-functions. This is because the degenerate spectral lines at
$n\,\hbar\,\omega_0$, for a fixed $n$, are all shifted by the same amount in the narrow wave-function regime, while
for wider wave-functions this degeneracy is removed, yielding a characteristic effect. The relative size of this effect
is comparable to the dynamical frequency shift, providing a second possible basis for an experimental test
of the \sne. We propose a particular experiment based on this gravitational fine-structure
in reference~\cite{Grossardt:2015c}.

It is also worth to remark that both effects, the dynamical and the spectral effect, scale proportional to the atomic
mass and the inverse cubed localisation of the atoms. This scaling factor is maximal for osmium, although
experimental requirements might necessitate a trade-off with other desirable properties.

A situation that has not been considered here, but might be of relevance for experimental tests of the \sne,
is self-gravitation of a superposition of (a small number of) energy-eigenstates.
A naive perturbative approach fails for times that are large compared to the oscillation period of the trap.
Hence, alternative approximation schemes are necessary in order to describe these states, that are neither
stationary nor Gaussian.

\section*{Acknowledgements}
The authors gratefully acknowledge funding and support from the John Templeton foundation (grant 39530).
AG acknowledges funding from the German Research Foundation (DFG). JB and HU acknowledge support from the
UK funding body {\small{EPSRC}} (EP/J014664/1). HU acknowledges financial support from the Foundational Questions
Institute (FQXi). AB acknowledges financial support from the EU project {\small{NANOQUESTFIT}}, {\small{INFN}}, and
the University of Trieste (grant FRA\,2013).

\appendix
\section*{Appendix}

\subsection{Axially symmetric stationary states}\label{app:axial}
In the discussion above we were, for reasons of simplicity, restricted to the one-dimensional \sne\
obtained in section~\ref{sec:reduction}. Realistic experimental scenarios
may however require---for practical reasons---that the assumption of
a strongly trapped wave-function in two dimensions must be given up. As a generalisation, here we discuss
the axially symmetric situation of a microsphere in a trap with frequency $\omega_0$ in $x$-direction, as
before, but a finite frequency $\mu\,\omega_0$ in $y$- and $z$-direction. We further assume, that the system
is in the ground state in $y$- and $z$-direction, such that the unperturbed state is
\begin{equation}\label{eqn:harm-osci-states-axial}
\psi^{(0)}_{n}(\vec r) = \frac{\sqrt{\mu}}{\sqrt{2^n \, n!}} \left( \frac{m\,\omega_0}{\pi\,\hbar} \right)^{3/4}
\exp \left[-\frac{m\,\omega_0}{2 \hbar}\,\left(x^2 + \mu(y^2 + z^2)\right)\right] 
H_n\left( \sqrt{\frac{m\,\omega_0}{\hbar}}\,x \right)\,,
\end{equation}
with the Hermite polynomials as defined in equation~\eqref{eqn:hermite}.

In full analogy to the one-dimensional derivation, we obtain the energy correction
\begin{align}\label{eqn:energy-correction-axial}
\Delta E_n &= -\frac{G\,\mu^2}{(2^n \, n!)^2} \, \left(\frac{m\,\omega_0}{\pi\,\hbar}\right)^3 \,
\int \D^3 r \, \int \D^3 r' \, H_n\left( \sqrt{\frac{m\,\omega_0}{\hbar}}\,x \right)^2 \,
H_n\left( \sqrt{\frac{m\,\omega_0}{\hbar}}\,x' \right)^2 \nnl
&\bleq \times \exp \left[-\frac{m\,\omega_0}{\hbar}\,\left(x^2 + x'^2 + \mu(y^2 + y'^2 + z^2 + z'^2)\right)\right] 
\,  I_{\rho_c}(\abs{\vec r - \vec r'}) \,.
\end{align}
Again, we introduce dimensionless variables
\begin{equation}
\xi = \sqrt{\frac{m\,\omega_0}{\hbar}}\,x \,, \quad\quad
s = \sqrt{\frac{m\,\omega_0}{\hbar}\,(y^2+z^2)} \,, \quad\quad
\alpha = 2\,\sigma\,\sqrt{\frac{m\,\omega_0}{\hbar}} \,, \quad\quad
\varrho = \frac{R}{\sigma} \,,
\end{equation}
as well as
\begin{equation}
\zeta = \frac{\abs{\vec r - \vec r'}}{2\,\sigma}
= \frac{1}{\alpha}\,\sqrt{(\xi-\xi')^2 + s^2 + {s'}^2 -2\,s\,s'\,\cos \varphi}\,.
\end{equation}
With this we get
\begin{subequations}\begin{align}\label{eqn:energy-correction-simplified-axial}
 \Delta E_n &= -\frac{G\,\hbar\,m_\text{atom}}{4\,\sigma^3\,\omega_0} \, f_n(\alpha,\,\varrho)
\intertext{with}
f_n(\alpha,\,\varrho) &= \frac{2\,\alpha^2\,\mu^2}{(2^n \, n!)^2\,\pi^2} \,
\int_{-\infty}^\infty \D \xi \, \int_{-\infty}^\infty \D \xi' \, \int_0^\infty \D s \, \int_0^\infty \D s'
 \, \int_0^{2\pi} \D \varphi \, \nnl
&\bleq \times s\,s' \, H_n\left( \xi \right)^2 \, H_n\left( \xi' \right)^2
\exp \left(-\xi^2 - \xi'^2 - \mu\,s^2 - \mu\, s'^2\right) 
\, i(\zeta,\,\varrho)  \,. \label{eqn:energy-correction-fn-axial}
\end{align}\end{subequations}
The difference to the previous form~\eqref{eqn:energy-correction} in the one-dimensional case is only in
the more complicated form, and $\mu$-dependence, of the integral function $f_n$.
This $f_n$ can not be solved analytically any more. Values can, however, still be obtained by numerical
integration.

\begin{figure}
\centering
\includegraphics[scale=.6]{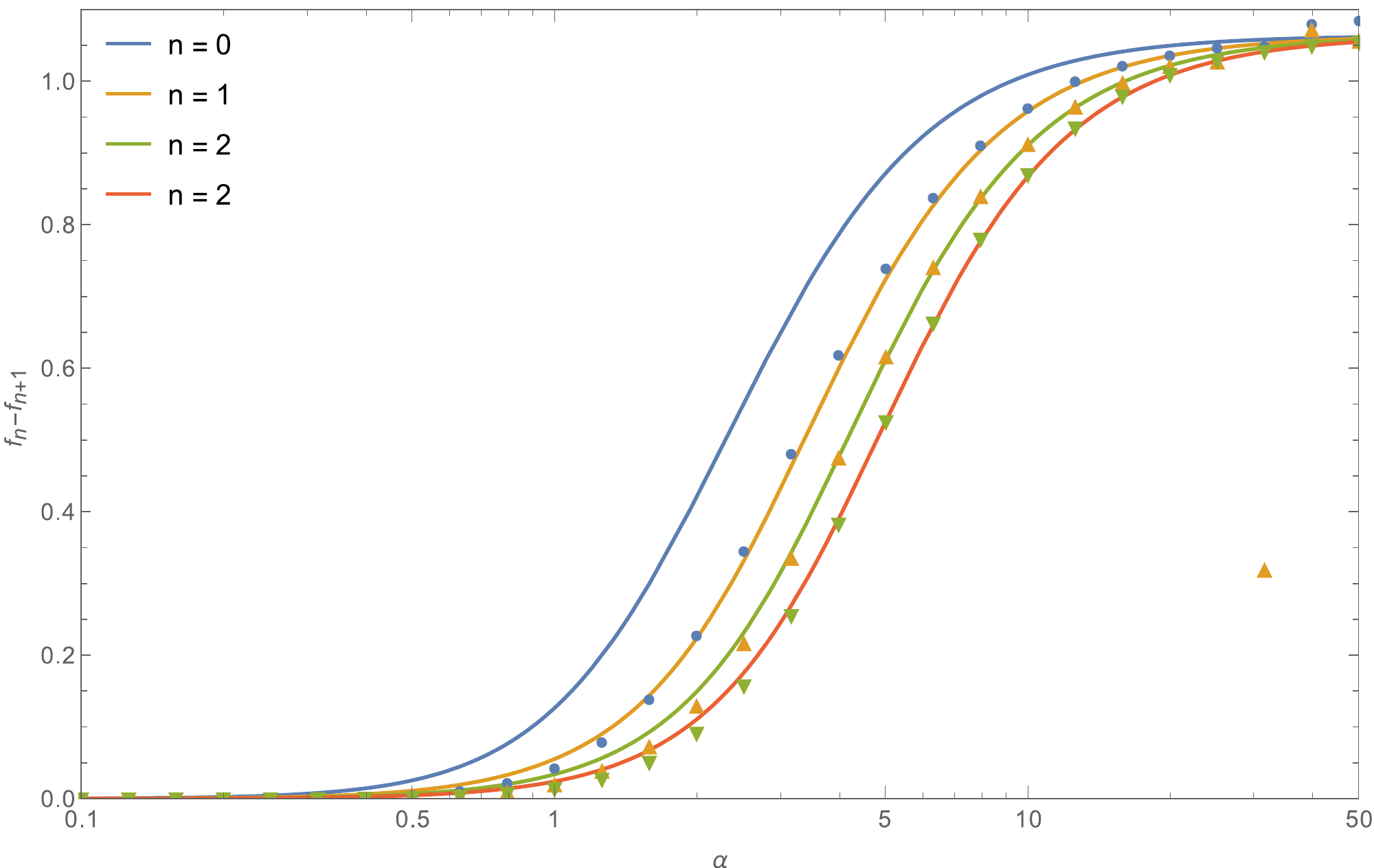}
\caption{Comparison of the values for $f_n - f_{n+1}$ for the analytically obtained one-dimensional
states (solid lines) and numerical results (data points) for the three-dimensional case with trap frequencies
$\omega_x = \omega_0 = 2 \omega_y = 2 \omega_z$. Same colours belong to same $n$.\label{fig:fn_num}}
\end{figure}

We are interested in transition energies in the intermediate regime for the Gaussian mass distribution, where
we---following the discussion in section~\ref{sec:spectrum-intermediate}---approximately take $\varrho \to \infty$.
Therefore we have
\begin{equation}
i(\zeta) \approx \frac{\erf\left(\sqrt{2}\,\zeta\right)}{2\,\zeta} \,
\end{equation}
where we neglected the constant term $\sim \beta_0$, because it does not contribute to transition energies,
and neglected the terms $\sim \beta_k$ for $k \leq 2$ because they are small.

The integral in equation~\eqref{eqn:energy-correction-fn-axial} can be further simplified for numerical evaluation
by substituting $u = \exp(-\mu\,s^2)$, and accordingly for $s'$. We then get
\begin{subequations}\begin{align}
f_n(\alpha) &= \frac{1}{2}\,\left(\frac{\alpha}{2^n \, n!\,\pi}\right)^2 \,
\int_{-\infty}^\infty \D \xi \, \int_{-\infty}^\infty \D \xi' \, \int_0^1 \D u \, \int_0^1 \D u'
 \, \int_0^{2\pi} \D \varphi \, \nnl
&\bleq \times H_n\left( \xi \right)^2 \, H_n\left( \xi' \right)^2
\exp \left(-\xi^2 - \xi'^2\right) 
\, \frac{\erf\left(\sqrt{2}\,\zeta\right)}{2\,\zeta} \,, \label{eqn:energy-correction-fn-axial-log}
\intertext{with}
\zeta &= \frac{1}{\alpha}\,\sqrt{(\xi-\xi')^2 -\frac{\ln u}{\mu} -\frac{\ln u'}{\mu}
-\frac{2}{\mu}\,\sqrt{\ln u \, \ln u'}\,\cos \varphi}\,.
\end{align}\end{subequations}
Since the integrand is highly oscillating, one must carefully choose a convenient numerical
integration method. We used the \emph{Divonne} algorithm from the \emph{Cuba} library~\cite{Hahn:2005}.

In figure~\ref{fig:fn_num} the numerically obtained results $f_n - f_{n+1}$ for a value $\mu = 1/2$ are plotted.
One can see that the effect discussed in section~\ref{sec:spectrum-intermediate} remains present also in this
fully three-dimensional situation, qualitatively and from its order of magnitude. Interestingly, the numerical
results suggest that for a trap frequency ratio of $2^k : 1 : 1$ the transition energies simply shift by $k$,
i.\,e. the transition $0 \to 1$ corresponds to the transition $k \to k+1$ in the one-dimensional case, and so on.
However, an analytical argument for this behaviour has yet to be found.

\subsection{Simplification of the function \boldmath$h(t)$}\label{app:function-h}
We want to show that the function $h(t)$ defined in~\eqref{eqn:function-h} can be written
\begin{equation}
h(t) = -\frac{1}{m^2}\,\left\langle p \, \frac{\partial V_g}{\partial x}
+ \frac{\partial V_g}{\partial x}\,p \right\rangle
= -\frac{2}{m}\,\frac{\partial}{\partial t}\,\langle V_g \rangle \,.
\end{equation}
First note that with the probability current density,
\begin{equation}
j(x) = \frac{\rmi\,\hbar}{2\,m}\,\left(\psi(x)\,\frac{\partial \psi^*(x)}{\partial x}
- \psi^*(x)\,\frac{\partial \psi(x)}{\partial x}\right) \,,
\end{equation}
we can write
\begin{equation}
\psi^*(x)\,\frac{\partial \psi(x)}{\partial x} = \frac{1}{2}\,\frac{\partial}{\partial x}\,\abs{\psi(x)}^2
+ \frac{\rmi\,m}{\hbar}\,j(x) \,.
\end{equation}
With this and the definition~\eqref{eqn:vg-one-dim} for the gravitational potential, we can write
\begin{align}
h(t) &= -\frac{\rmi\,\hbar\,G}{m^2}\,\int \D x \, \int \D x' \,\Bigg( \abs{\psi(x)}^2\,\abs{\psi(x')}^2
\,\frac{\partial^2 I_{\rho_c}(\abs{x-x'})}{\partial x^2} \nnl
&\bleq + 2\,\psi^*(x)\,\frac{\partial \psi(x)}{\partial x}
\,\abs{\psi(x')}^2\,\frac{\partial I_{\rho_c}(\abs{x-x'})}{\partial x}\Bigg) \\
&= -\frac{\rmi\hbar\,G}{m^2}\int \D x'\abs{\psi(x')}^2\int \D x \Bigg( \abs{\psi(x)}^2
\frac{\partial^2 I_{\rho_c}(\abs{x-x'})}{\partial x^2} + \frac{\partial \abs{\psi(x)}^2}{\partial x}
\,\frac{\partial I_{\rho_c}(\abs{x-x'})}{\partial x}\Bigg) \nnl
&\bleq +\frac{2\,G}{m} \, \int \D x'\,\abs{\psi(x')}^2\,\int \D x\,
j(x)\,\frac{\partial I_{\rho_c}(\abs{x-x'})}{\partial x} \\
&= -\frac{2\,G}{m} \, \int \D x'\,\abs{\psi(x')}^2\,\int \D x\,
\frac{\partial j(x)}{\partial x}\,I_{\rho_c}(\abs{x-x'}) \\
&= \frac{2\,G}{m} \, \int \D x'\,\abs{\psi(x')}^2\,\int \D x\,
\frac{\partial \abs{\psi(x)}^2}{\partial t}\,I_{\rho_c}(\abs{x-x'}) \\
&= \frac{G}{m} \, \int \D x'\,\int \D x\,
\frac{\partial}{\partial t}\left(\abs{\psi(x')}^2\,\abs{\psi(x)}^2\right)\,I_{\rho_c}(\abs{x-x'}) \\
&= -\frac{1}{m} \, \frac{\partial}{\partial t} \, \left\langle V_g \right\rangle \,,
\end{align}
where we made use of the continuity equation in the second to last step.

\end{document}